\documentclass[10pt,journal,comsoc]{IEEEtran}
\usepackage{graphicx} 
\usepackage{graphics}
\usepackage{multirow}
\usepackage{textcomp}
\usepackage{array}
\usepackage{xcolor}

\usepackage{tabularx}
\newcolumntype{Y}{>{\centering\arraybackslash}X} 

\usepackage[caption=false,font=footnotesize]{subfig}

\usepackage{tikz}
\usetikzlibrary{shapes.geometric, arrows, positioning, calc}

\usepackage{amsmath,amssymb,amsfonts}
\usepackage{bm}

\usepackage{algorithm}
\usepackage{algorithmic}
\usepackage[algo2e,linesnumbered,ruled]{algorithm2e}

\usepackage{amsthm}
\theoremstyle{definition}

\newtheorem{definition}{Definition}

\newtheorem{assumption}{Assumption}

\SetKwProg{Fn}{Function}{}{end}

\SetAlgoSkip{}

\newcolumntype{M}[1]{>{\centering\arraybackslash}m{#1}}

\DeclareSymbolFont{symbolsC}{U}{txsyc}{m}{n}
\DeclareMathSymbol{\notniFromTxfonts}{\mathrel}{symbolsC}{61}

\title{Adaptive Probabilistic Skyline Analytics for Mobile Edge-Cloud Systems via State-Aware Deep Reinforcement Learning}

\author{Chuan-Chi~Lai,~\IEEEmembership{Member,~IEEE}
    \IEEEcompsocitemizethanks{
        \IEEEcompsocthanksitem{This research was supported by the National Science and Technology Council, Taiwan, under Grant Nos. NSTC 114-2221-E-194-062- and NSTC 115-2221-E-194-042-MY2. 
		This work was also partially supported by the Advanced Institute of Manufacturing with High-tech Innovations (AIM-HI) from the Featured Areas Research Center Program within the framework of the Higher Education Sprout Project by the Ministry of Education (MOE) in Taiwan. \emph{(Corresponding author: Chuan-Chi~Lai.)}}
        \IEEEcompsocthanksitem{C.-C. Lai is with the Department of Communications Engineering, National Chung Cheng University, Minxiong Township, Chiayi County 621301, Taiwan, and also with the Advanced Institute of Manufacturing with High-tech Innovations (AIM-HI), National Chung Cheng University, Minxiong Township, Chiayi County 621301, Taiwan (e-mail: chuanclai@ccu.edu.tw).}
    }
}

\IEEEtitleabstractindextext{%
\begin{abstract}
The proliferation of Mobile Edge Computing (MEC) and distributed sensing necessitates efficient Probabilistic Skyline (PSKY) query analytics at the network edge. However, this process is severely constrained by the computation-communication trade-off under constrained wireless capacity and dynamic traffic conditions. Furthermore, mobility-induced workload fluctuations and spatio-temporal data shifts render static filtering policies inefficient, often triggering network congestion or compromising query fidelity. To address these systemic inefficiencies in dynamic mobile environments, this paper introduces SA-PSKY, a self-adaptive framework integrating deep reinforcement learning into a distributed query optimization architecture. We formulate the probabilistic skyline filtering as an adaptive control problem and develop a State-Aware Adaptive Weighting (SAAW) mechanism to dynamically regulate computation, communication, and fidelity. By incorporating Prioritized Experience Replay (PER) to stabilize learning around abrupt network transitions, our framework reliably navigates the Pareto-efficient operating point. Empirical evaluations confirm that SA-PSKY achieves substantial latency reduction while preserving query fidelity under constrained communication resources. Furthermore, zero-shot robustness analyses reveal superior adaptation to unseen mobility-induced uncertainty shifts, where rigid methods exhibit severe policy degradation.
\end{abstract}

}

\begin{document}



\maketitle
\IEEEdisplaynontitleabstractindextext

%
\IEEEpeerreviewmaketitle

\section{Introduction}
\label{sec:intro}

\IEEEPARstart{I}{n} the contemporary \textit{Mobile Edge Computing} (MEC) landscape, massive numbers of mobile devices and sensors, such as connected vehicles and mobile crowdsensing nodes, continuously generate vast amounts of spatio-temporally varying uncertain data \cite{11244906,11074426,10545344}. Among various data analytics techniques, \textit{Probabilistic Skyline} (PSKY) queries have emerged as a critical tool for multi-criteria decision-making under uncertainty \cite{10.5555/1325851.1325858}. By identifying objects that are likely to be optimal across conflicting attributes despite stochastic noise \cite{11098465}, PSKY analytics empower intelligent services at the network edge.

However, deploying PSKY queries within a mobile edge-cloud architecture introduces significant challenges. To alleviate the bottleneck of transmitting raw data over bandwidth-constrained wireless links, edge nodes typically apply a local filtering threshold $\alpha$ to prune unpromising candidates before sending them to a central cloud broker \cite{8731646,10.1007/s00778-009-0162-1}. While fixed thresholding strategies may suffice in static environments, they become highly inefficient in mobile environments. The inherent mobility of users and sensing sources induces rapid distribution shifts and severe workload fluctuations. Consequently, the spatial candidate density and the resulting wireless traffic load vary drastically over short temporal windows \cite{10595132,10175534}.

This volatility exposes a fundamental tension: the non-linear computation-communication-fidelity trade-off. Applying a stricter filtering threshold proactively prunes the search space, significantly reducing both the local computational expenditure at the edge and the communication payload over the wireless channel \cite{9374102}. Yet, it elevates the risk of erroneously pruning true global skyline objects, thereby severely reducing query fidelity. Conversely, relaxed filtering preserves query fidelity but overwhelms the edge processors with $O(N^2 \cdot m^2 \cdot d)$ exhaustive instance-level dominance checks \cite{8302507} and saturates the shared wireless bandwidth, leading to severe network congestion and broker queue overflow \cite{10298037,9633191,9354847}.

Therefore, the traditional paradigm of static or heuristic-based thresholding is fundamentally inadequate. The overarching research challenge can be formalized by a central question: \textit{How can a mobile edge-cloud system dynamically regulate local probabilistic skyline filtering under mobility-induced workload and bandwidth variations while preserving query fidelity?} To solve this adaptive computation-communication-fidelity control problem, advanced optimization frameworks are required \cite{9944146}.

To directly address this challenge, we propose the \textit{Self-Adaptive Probabilistic Skyline Analytics} (SA-PSKY) framework. By formulating the distributed edge-filtering operation as a continuous-space \textit{Markov Decision Process} (MDP), SA-PSKY utilizes \textit{Deep Reinforcement Learning} (DRL) to learn an adaptive Pareto-efficient operating point under mobile resource dynamics \cite{10597395}. Rather than relying on rigid parameter configurations, our framework intelligently anticipates the multi-dimensional impact of threshold adjustments under mobility-induced shifts, bridging the gap between local resource constraints and global system responsiveness. To ensure learning stability, we incorporate \textit{Prioritized Experience Replay} (PER)~\cite{per2016} as a supporting mechanism for the proposed controller. 

The principal technical contributions of this research are summarized as follows:

\begin{itemize}
    \item \textbf{Mobile Query-Control Formulation}: We rigorously formulate probabilistic skyline filtering as a closed-loop adaptive control problem under mobility-induced workload and communication dynamics. This formulation explicitly captures the non-linear interdependencies among stochastic data arrivals, heterogeneous uncertainty distributions, and strict wireless bandwidth limits.
    
    \item \textbf{State-Aware Adaptive Filtering}: We design a State-Aware Adaptive Weighting (SAAW) mechanism. By leveraging mobile network queue pressure as a dynamic control state, the filtering policy autonomously transitions to rebalance optimization priorities among local computational burdens, global transmission overhead, and query fidelity.
    
    \item \textbf{Zero-Shot Mobility Robustness}: Through comprehensive empirical evaluations, we demonstrate that SA-PSKY exhibits strong zero-shot generalization performance. It autonomously adapts to unseen mobility-induced uncertainty shifts without requiring policy retraining, providing a practical framework for real-time analytics in large-scale mobile edge ecosystems.
\end{itemize}

\section{Related Work}
\label{sec:related_work}

Recent advances in mobile edge computing have shifted distributed data analytics toward collaborative edge-cloud architectures that can adapt to dynamic workloads and resource conditions \cite{11074426}. This section provides a systematic review of the literature, focusing on the theoretical foundations of skyline queries, distributed mobile processing architectures, adaptive resource management for mobile edge systems, and deep reinforcement learning for edge query optimization.

\subsection{Foundations of Skyline Queries over Uncertain Data}

The skyline operator was originally introduced as a method for identifying a set of non-dominated objects based on multiple criteria \cite{914855}. In the domain of uncertain data, the PSKY query was developed to handle stochastic attribute values by assigning an existence probability to each object within the skyline set \cite{10.5555/1325851.1325858}. Early research in this field focused on improving the efficiency of dominance checks through various indexing structures, such as Z-order curves and R-tree variations \cite{5394984,4221738}. To address the challenges posed by continuous data streams, subsequent studies introduced the concept of the sliding window PSKY, which allows for real-time updates of the skyline set as new objects arrive \cite{1410162,LIU201540}. Further refinements led to the development of the $k$-dominant skyline and the reverse skyline query, both of which offer more flexible filtering criteria for complex decision-making scenarios \cite{10298037,10.1145/1670243.1670246}. 

Most existing studies focus on improving the computational efficiency of skyline evaluation, while the filtering strategy is typically treated as a predefined query or system parameter. Adaptive threshold optimization for probabilistic skyline processing in dynamic mobile edge environments remains comparatively unexplored.

\subsection{Distributed and Parallel Processing Architectures}

To handle the massive data volumes generated by distributed mobile sensors, research has increasingly focused on parallelizing and distributing the skyline operator across network nodes \cite{9714833,9186333}. Distributed skyline processing typically involves two stages: local pruning at the data sources and global aggregation at a central coordinator \cite{8731646}. Various partitioning strategies, including grid-based and angle-based partitioning, have been proposed to balance the workload across multiple nodes \cite{7379908,10.1145/1376616.1376642}. In large-scale network environments, \textit{peer-to-peer} (P2P) architectures were explored to eliminate the single point of failure inherent in centralized brokers \cite{4731252,10.1007/11687238_10}. 

For wireless sensor and mobile networks, specific space-efficient algorithms were developed to minimize the memory footprint on nodes while maintaining high query accuracy \cite{SU2010680}. Several frameworks \cite{9714833,9186333} have utilized the MapReduce paradigm and Spark-based implementations to achieve scalability in big data environments, though these methods often introduce significant communication latencies that are unsuitable for latency-sensitive mobile edge tasks \cite{10316604}. However, these approaches primarily improve parallel execution efficiency while generally assuming static filtering policies rather than adaptive threshold control under mobility-induced network fluctuations.

\subsection{Adaptive Resource Management and Edge Computing}

The emergence of mobile edge computing has necessitated a more dynamic approach to resource management \cite{9354847,10356753}. Early adaptive systems relied on control theory or heuristic rules to adjust system parameters such as sampling rates and transmission frequencies \cite{7488250}. Lyapunov optimization techniques were later introduced to provide theoretical guarantees for system stability and energy efficiency under stochastic workloads \cite{2010Neely}. In the context of data filtering, some studies proposed rule-based adaptive thresholds that respond to changes in network congestion or CPU utilization \cite{NGUYEN2024103884}. However, these traditional adaptive methods often struggle with the non-linear and high-dimensional nature of modern mobile data streams. Comprehensive surveys on edge-assisted data management highlight that static or simplistic adaptive strategies cannot effectively navigate the complex trade-offs between local computation and global communication in heterogeneous mobile environments \cite{9863881}. 

In this regard, recent efforts have demonstrated the effectiveness of Markov-inspired optimization frameworks for managing dynamic system state transitions, enabling adaptive task scheduling and resource allocation in end-edge-cloud environments \cite{11494288}. Nevertheless, these studies primarily optimize generic computing, communication, or task-level resources rather than query-level filtering decisions for probabilistic skyline analytics.

\subsection{Deep Reinforcement Learning for Query Optimization}

DRL has recently been recognized as a powerful tool for optimizing data management systems. By leveraging the representational power of deep neural networks, DRL agents can learn effective control policies directly from observed system states. In the field of query optimization, DRL has been applied to tasks such as join order selection and index tuning \cite{9723570}. The transition from discrete models, like \textit{Deep Q-Networks} (DQN) \cite{mnih2015human}, to continuous control algorithms, such as \textit{Deep Deterministic Policy Gradient} (DDPG) \cite{lillicrap2016}, has enabled more precise adjustments of system parameters. Recent research has demonstrated that DRL has been shown to be effective for task offloading, bandwidth allocation, and collaborative resource management in distributed edge computing environments \cite{9944146}.

Furthermore, latest advancements in query processing highlight the ongoing necessity for sophisticated optimization techniques to handle complex constraints in probabilistic environments \cite{Gao2025}. Overall, existing studies reveal three distinct gaps: (1) Existing PSKY methods optimize query evaluation efficiency but assume predefined filtering policies; (2) Existing MEC frameworks optimize generic system resources or offloading targets rather than query-level semantics; (3) Existing DRL-based edge optimization methods generally focus on generic control decisions rather than probabilistic skyline filtering.

What remains missing is a query-semantic control loop that directly maps mobile system states to probabilistic skyline filtering decisions while explicitly accounting for recall loss. This gap motivates the proposed SA-PSKY framework, which formulates skyline threshold optimization as a continuous-control MDP to address the mobility-induced computation-communication-fidelity trade-off.

\section{System Model and Problem Definition}
\label{sec:system_model}

In this section, we establish the theoretical foundations for probabilistic skyline analytics and formalize the joint optimization problem. We model the complex interplay between local edge computation and global cloud communication using queuing theory and computational complexity analysis. To ensure mathematical tractability, the notations used throughout this paper are summarized in Table~\ref{tab:notations}. We first introduce the necessary definitions and system assumptions.

\subsection{Mobile Sensing and Uncertain Data Streams}

Consider a distributed MEC environment where a set of mobile sensing nodes, such as connected vehicles or mobile crowdsensing devices, denoted as $\mathcal{E} = \{e_1, e_2, \dots, e_K\}$, continuously collect sensor readings from the physical world. Due to unavoidable sensing noise, transmission errors, and privacy-preserving perturbations, the collected data is inherently uncertain.

We model each uncertain object $u^{(i)}$ generated at a mobile edge node $e_i$ as a set of discrete instances $u^{(i)} = \{u^{(i)}_1, u^{(i)}_2, \dots, u^{(i)}_m\}$. Here, each instance $u^{(i)}_j \in \mathbb{R}^d$ represents a possible state of the object in a $d$-dimensional feature space, associated with an existence probability $P(u^{(i)}_j)$. To ensure statistical consistency within the probability space, we enforce the following normalization constraint:
\begin{equation}
    \sum_{j=1}^{m} P(u^{(i)}_j) \leq 1,
\end{equation}
where the inequality allows for the possibility that the object may not exist at all.

\begin{assumption}[Statistical Independence]
    We assume that the uncertain objects generated by different mobile sensors are statistically independent. This assumption implies that the probability distributions of any two distinct objects $u^{(i)}$ and $u^{(k)}$ are uncorrelated, allowing their joint probability to be calculated as the product of their individual marginal probabilities.
\end{assumption}

In practical mobile scenarios, sensor data is not static but arrives sequentially, forming an unbounded stream.

\begin{definition}[Uncertain Data Stream]
    In the dynamic mobile environment, the raw sensing data generated by an edge node $e_i$ forms an unbounded uncertain data stream, denoted as $\mathcal{R}_i$.
    \begin{equation}
        \mathcal{R}_i = \langle u^{(1)}, u^{(2)}, \dots, u^{(t)}, \dots \rangle,
    \end{equation}
    where each $u^{(t)}$ represents an uncertain object arriving at a discrete time step $t$. The stream is strictly ordered by arrival time, and the volume of data is assumed to be infinite over the system's operational lifespan.
\end{definition}

\begin{table}[!t]
    \caption{Summary of Key Notations for System Model and Agent Formulation}
    \label{tab:notations}
    \centering
    \begin{tabularx}{\columnwidth}{|c|X|}
    \hline
    \textbf{Symbol} & \textbf{Description} \\ 
    \hline
    \hline
    $\mathcal{E}$ & Set of edge nodes $\{e_1, \dots, e_K\}$ \\
    \hline
    $K$ & Total number of edge nodes in the system \\ 
    \hline
    $\mathcal{R}_i$ & Uncertain data stream generated at node $e_i$ \\ 
    \hline
    $u$ & An uncertain object \\
    \hline
    $m$ & Number of instances per uncertain object \\
    \hline
    $d$ & Number of dimensions in the feature space \\ 
    \hline
    $\mathcal{W}_i$ & Sliding window set at node $e_i$ \\
    \hline
    $\mathcal{D}_i$ & Active local dataset at node $e_i$ \\ 
    \hline
    $W_{\max}$ & Maximum capacity of the sliding window \\
    \hline
    $N$ & Current number of active objects in the window \\
    \hline
    $N_{\alpha}$ & Number of candidate objects satisfying the local threshold \\
    \hline
    $\mathcal{S}_i$ & Filtered candidate set transmitted from node $e_i$ \\ 
    \hline
    $P_{\rm sky}(u)$ & Global skyline probability of object $u$ \\
    \hline
    $P_{\rm local}(u)$ & Local skyline probability of object $u$ computed at the edge \\ 
    \hline
    $\alpha_{i,t}$ & Local filtering threshold of node $e_i$ at time $t$ \\
    \hline
    $\boldsymbol{\alpha}_t$ & Joint action vector of thresholds at time $t$ \\ 
    \hline
    $\Phi(\alpha)$ & Pruning efficiency factor for local dominance checks \\
    \hline
    $\sigma_i(\alpha_{i,t})$ & Local selectivity ratio at node $e_i$ at time $t$ \\
    \hline
    $c_{\rm comp}$ & Calibrated unit computation cost per retained candidate \\
    \hline
    $c_{\rm trans}$ & Unit transmission cost per object over the uplink channel \\
    \hline
    $T_{\rm comp}$ & Local computation time at the edge node \\
    \hline
    $T_{\rm trans}$ & Data transmission time over the uplink channel \\ 
    \hline
    $T_{\rm cloud}$ & Cloud processing and queuing time \\
    \hline
    $L_{\rm sys}(t)$ & Absolute end-to-end system latency at time $t$ \\ 
    \hline
    $\lambda_{i,t}$ & Dynamic local data arrival rate at node $e_i$ at time $t$ \\
    \hline
    $\Lambda_t(\boldsymbol{\alpha}_t)$ & Dynamic aggregate data arrival rate at the cloud broker \\
    \hline
    $\mu$ & Service rate of the cloud broker \\ 
    \hline
    $\rho_t$ & Traffic intensity ($\Lambda_t(\boldsymbol{\alpha}_t) / \mu$) at time $t$ \\
    \hline
    $B$ & Bandwidth of the shared uplink channel \\ 
    \hline
    $U_{\rm sys}(t)$ & Net physical system utility at time $t$ \\
    \hline
    $s_t$ & State vector observed by the agent at time $t$ \\
    \hline
    $Q_t$ & Instantaneous broker queue length \\
    \hline
    $\boldsymbol{\lambda}_t$ & Vector of normalized data arrival rates from edge nodes \\
    \hline
    $\mathcal{D}_t$ & Spatial distribution feature vector at time $t$ \\
    \hline
    $\mathbf{c}_{i,t}, \mathbf{g}_t$ & Local feature centroid and global virtual center \\
    \hline
    $\mathbf{G}_t, \mathbf{V}_t$ & Centroid distance profile and correlation metric vector \\
    \hline
    $a_t$ & Action vector generated by the agent at time $t$ \\
    \hline
    $r_t$ & Shaped instantaneous reward at time $t$ \\
    \hline
    $C_{\max}, L_{\max}$ & Normalization constants for aggregate cost and latency \\
    \hline
    $\tilde{\mathcal{U}}_{t}$ & Normalized data utility (system hit rate) \\
    \hline
    $\mathcal{P}_t$ & Filtering precision of the transmitted candidates \\ 
    \hline
    $\Omega_{t}$ & Global data selectivity ratio \\
    \hline
    $\bar{\epsilon}_{t}$ & Average false negative rate across all edge nodes \\ 
    \hline
    $w_i$ & State-aware adaptive weighting coefficients ($i \in \{1,2,3,4\}$) \\ 
    \hline
    \end{tabularx}
\end{table}

\subsection{Mobility-Induced System Dynamics}

Unlike static networks, the physical mobility of sensing nodes introduces severe spatio-temporal variations into the system. Let $\mathbf{l}_i(t)$ denote the geographical location of a mobile node $e_i$ at time $t$. The node's mobility trajectory dictates its sensing context, directly impacting the fundamental variables of the distributed filtering system. This mobility-induced dynamic forms a strict causal chain that reshapes the skyline analytics:

\begin{itemize}
    \item \textbf{Time-Varying Arrival Rates}: The local data generation rate $\lambda_{i,t}$ at node $e_i$ becomes a highly dynamic function of its mobility. For instance, a mobile sensing node entering a high-density urban hotspot will encounter a sudden burst of sensing events, causing $\lambda_{i,t}$ to spike unpredictably.
    \item \textbf{Spatial Distribution Shifts}: As mobile nodes traverse different physical regions, the underlying multi-dimensional features of the collected data evolve. We abstract this behavior using a spatial distribution feature $\mathcal{D}_t = f(\mathbf{l}_i(t), \mathbf{v}_i(t), \varrho_i(t))$, where $\mathbf{v}_i(t)$ and $\varrho_i(t)$ denote the node's velocity vector and local spatial density, respectively. This mobility-driven shift continuously alters the mutual dominance probabilities among uncertain objects, drastically changing the local candidate survival density.
    \item \textbf{Fluctuating Communication Load}: Consequently, the aggregate network traffic and the global selectivity ratio become inherently non-stationary. The dynamic candidate density forces the shared wireless channel into unpredictable states of congestion.
\end{itemize}

This causal chain demonstrates that mobility is the root cause of workload and bandwidth variations. It establishes the explicit necessity for an adaptive control mechanism, as a static filtering threshold cannot simultaneously mitigate the fluctuating computational load and communication delay.

\subsection{Sliding Window Model and Probabilistic Dominance}

To continuously process the unbounded data streams and evaluate the corresponding skyline probabilities, we employ a sliding window mechanism coupled with a probabilistic dominance framework. We first define the local data management strategy at the edge.

\begin{definition}[Sliding Window Model]
    To process the unbounded stream $\mathcal{R}_i$ under finite memory and latency constraints, each edge node employs a count-based sliding window mechanism, denoted as $\mathcal{W}_i$. Let $W_{\max}$ be the maximum capacity of the window. At any given time $t$, the window $\mathcal{W}_i(t)$ maintains the most recent set of $N$ objects from the stream, where $N \le W_{\max}$.
    The window updates follow a first-in-first-out policy. Let $u_{\rm old}$ denote the oldest object in the window. The update rule is:
    \begin{equation}
        \mathcal{W}_i(t) = 
        \begin{cases} 
            \mathcal{W}_i(t-1) \cup \{u^{(t)}\}, & |\mathcal{W}_i| < W_{\max} \\
            (\mathcal{W}_i(t-1) \setminus \{u_{\rm old}\}) \cup \{u^{(t)}\}, & |\mathcal{W}_i| = W_{\max}
        \end{cases}
    \end{equation}
    Consequently, the local probabilistic skyline computation at time $t$ is performed strictly on the active dataset $\mathcal{D}_i(t) = \mathcal{W}_i(t)$.
\end{definition}

The core concept of skyline queries relies on the dominance relationship. In a deterministic setting, dominance is binary, allowing for straightforward comparisons. However, in our uncertain setting, the presence of multiple possible instances for each object necessitates a probabilistic framework to quantify the dominance likelihood. We begin by defining the dominance criteria at the instance level.

\begin{definition}[Instance-Level Dominance]
    Let $u_A$ and $u_B$ be two uncertain objects. Let $u_{A,p}$ and $u_{B,q}$ be specific instances of $u_A$ and $u_B$ respectively in the $d$-dimensional space. We say $u_{A,p}$ dominates $u_{B,q}$, denoted as $u_{A,p} \prec u_{B,q}$, if and only if $u_{A,p}$ is strictly better (smaller) in at least one dimension and not worse in all other dimensions:
    \begin{equation}
    \begin{cases} 
    \forall r \in [1,d], & v_{A,p,r} \leq v_{B,q,r} \\
    \exists r \in [1,d], & v_{A,p,r} < v_{B,q,r}
    \end{cases}
    \end{equation}
    where $v_{A,p,r}$ denotes the value of the $r$-th attribute of instance $u_{A,p}$.
\end{definition}

While instance-level dominance is deterministic, an uncertain object is essentially a probability distribution over its instances. Therefore, we extend this concept to quantify the likelihood that one uncertain object dominates another by aggregating the probabilities across their joint instance space.

\begin{definition}[Object-Level Dominance Probability]
    The probability that uncertain object $u_A$ dominates object $u_B$ is defined as the sum of the joint probabilities of all instance pairs where dominance holds:
    \begin{equation}
        P(u_A \prec u_B) = \sum_{u_{A,p} \in u_A} \sum_{u_{B,q} \in u_B} P(u_{A,p}) P(u_{B,q}) \cdot \mathbb{I}(u_{A,p} \prec u_{B,q}),
    \end{equation}
    where $\mathbb{I}(\cdot)$ is an indicator function that returns 1 if the dominance condition is met, and 0 otherwise.
\end{definition}

With the dominance probability established, we characterize an object's likelihood of belonging to the global skyline. An object is considered a candidate if it remains un-dominated by all other objects in the dataset with sufficient confidence.

\begin{definition}[Probabilistic Skyline]
    An object $u$ is considered a probabilistic skyline candidate if it has a sufficiently high probability of remaining un-dominated by the rest of the dataset. Formally, for a global dataset $\mathcal{U}$, the skyline probability of $u$ is given by:
    \begin{equation}\label{eq:skyline_probability}
        P_{\rm sky}(u) = \prod_{v \in \mathcal{U}, v \neq u} (1 - P(v \prec u)).
    \end{equation}
    If $P_{\rm sky}(u) \ge \alpha$, where $\alpha \in [0,1]$ is a filtering threshold, then $u$ is part of the $\alpha$-probabilistic skyline.
\end{definition}

\subsection{Distributed Edge-Cloud Architecture}
\begin{figure}[!t]
    \centering
    \includegraphics[width=\linewidth]{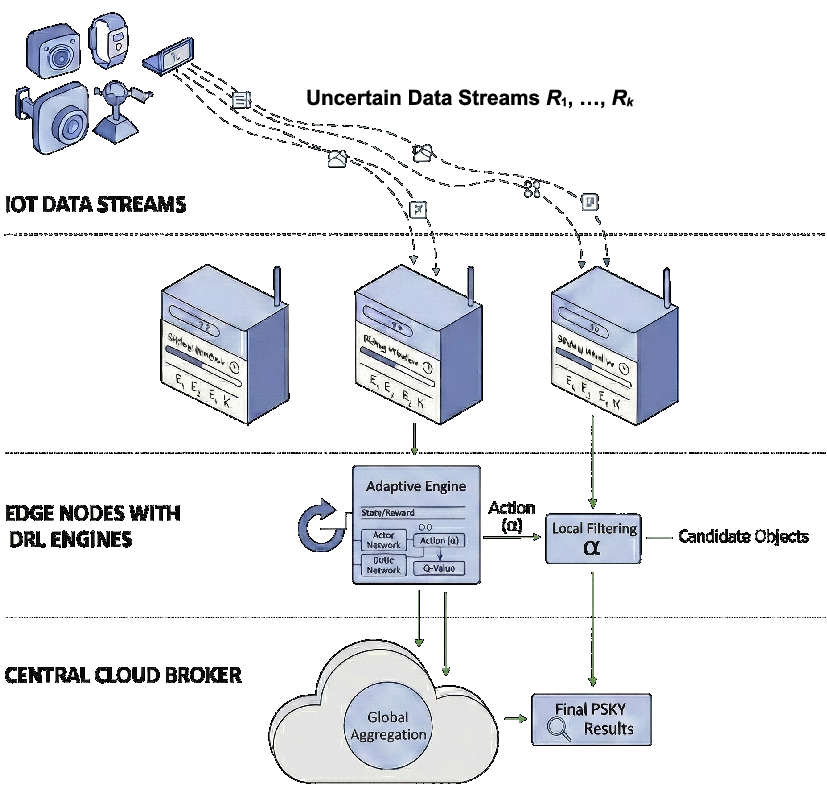}
    \caption{Distributed Edge-Cloud Architecture for Probabilistic Skyline Query Processing. Each edge node independently filters local data streams before transmitting candidates to a central broker for final aggregation.}
    \label{fig:system_model}
\end{figure}

The proposed framework operates on a hierarchical architecture, specifically engineered to optimize the resource-constrained interaction between the edge and cloud layers. As depicted in Fig.~\ref{fig:system_model}, the edge layer comprises heterogeneous nodes executing localized data ingestion and preliminary filtering, while the cloud layer functions as a centralized broker for global skyline aggregation.

\subsubsection{Edge Layer (Local Filtering)}
The edge layer consists of nodes with finite computational capacity. Given that each node $e_i$ only maintains a local data subset $\mathcal{D}_i \subset \mathcal{U}$, it cannot compute the true global skyline probability defined in \eqref{eq:skyline_probability}. Instead, it derives a local skyline probability $P_{\rm local}(u^{(i)})$ restricted to $\mathcal{D}_i$. Each node employs the sliding window $\mathcal{W}_i$ to manage streaming data and applies a dynamic filtering threshold $\alpha_{i,t}$. Objects failing the criterion $P_{\rm local}(u^{(i)}) \ge \alpha_{i,t}$ are pruned locally, resulting in the candidate set $\mathcal{S}_i$. Given $P_{\rm local}(u^{(i)}) \ge P_{\rm sky}(u^{(i)})$, under the exact-preserving condition $\alpha_{i,t} \le \alpha_{\rm query}$, local filtering does not discard any object satisfying the global query threshold.

\subsubsection{Cloud Layer (Global Aggregation)}
The cloud broker is responsible for global consistency. Upon receiving candidate sets $\bigcup_{i=1}^K \mathcal{S}_i$, the broker executes high-level dominance verifications to finalize the global skyline result.

\subsection{Computational Complexity Model}
Probabilistic skyline queries incur $O(N^2 \cdot m^2 \cdot d)$ brute-force complexity, rendering exhaustive instance-level dominance checks prohibitive in resource-constrained IoE environments. This significant computational burden necessitates an adaptive filtering mechanism to prune the search space and maintain real-time responsiveness within the edge-cloud continuum.

As an empirical surrogate model of local computation cost, the processing time is formulated as:
\begin{equation}\label{eq:comp_surrogate}
    T_{\rm comp}(i, \alpha_{i,t}) = c_{\rm comp} \cdot |\mathcal{S}_i(\alpha_{i,t})|,
\end{equation}
where $c_{\rm comp}$ represents the calibrated unit cost obtained from local dominance-processing profiling. This expression captures the dominant candidate-comparison workload after threshold-guided pruning and is used as a tractable surrogate for the physical execution time. The surrogate represents the post-pruning candidate-processing workload rather than the full raw dominance-evaluation cost; the latter is characterized separately by the $O(N^2 \cdot m^2 \cdot d)$ complexity analysis. Increasing the filtering threshold $\alpha_{i,t}$ imposes a stricter condition, which significantly reduces the selectivity ratio and consequently decreases $|\mathcal{S}_i(\alpha_{i,t})|$. This proactive reduction directly alleviates both the local computational load $T_{\rm comp}$ and the transmission overhead $T_{\rm trans}$. Navigating this inherent trade-off between local computational expenditure and global network efficiency constitutes the core optimization challenge in distributed PSKY analytics.

\subsection{Network Channel and Queuing Model}
\label{sec:communication_queuing}

In realistic mobile environments, edge nodes often share a bandwidth-constrained wireless uplink channel to communicate with the cloud. We model this channel contention and the subsequent processing at the broker using queuing theory.

\begin{assumption}[Poisson Arrivals and Queuing]
    We assume that the arrival of skyline candidates from the distributed edge nodes follows a Poisson process, and the service time at the broker follows an exponential distribution, allowing us to model the system as an M/M/1 queue. According to the Palm-Khintchine theorem, the superposition of packet traffic from a massive number of independent mobile nodes asymptotically approaches a Poisson process. Thus, the M/M/1 model provides a tractable approximation for the broker-side queuing behavior.
\end{assumption}

Let $\omega$ be the average size of a candidate object (in bits) and $B$ be the total shared bandwidth. We define the unit transmission cost as $c_{\rm trans} = (\omega \cdot \xi) / B$, where $\xi \ge 1$ is a communication overhead factor that accounts for packet headers, MAC layer signaling, and retransmission overheads in practical mobile edge deployments. The transmission time for node $i$ under threshold $\alpha_{i,t}$ is thus formulated as 
\begin{equation}
    T_{\rm trans}(i, \alpha_{i,t}) = |\mathcal{S}_i(\alpha_{i,t})| \cdot c_{\rm trans}.
\end{equation} 
To capture the mobility-induced workload variations, the aggregate arrival rate $\Lambda_t(\boldsymbol{\alpha}_t)$ at the cloud at time $t$ is derived as:
\begin{equation}
    \Lambda_t(\boldsymbol{\alpha}_t) = \sum_{i=1}^K \lambda_{i,t} \cdot \sigma_i(\alpha_{i,t}),
\end{equation}
where $\lambda_{i,t}$ is the dynamic raw data generation rate at mobile node $e_i$, and $\sigma_i(\alpha_{i,t}) \in [0,1]$ is the local selectivity ratio.

The cloud processing time $T_{\rm cloud}$, which includes both waiting time in the buffer and service time at the broker, is formulated as a piecewise function:
\begin{equation}
    T_{\rm cloud}(\rho_t) = 
    \begin{cases}
        \frac{1}{\mu - \Lambda_t(\boldsymbol{\alpha}_t)}, & \rho_t < 1 \\
        T_{\rm cloud}^{\max}, & \rho_t \ge 1
    \end{cases}
\end{equation}
where $\mu$ is the service rate of the broker. For $\rho_t \ge 1$, the broker enters an overflow state; the standard queuing expression is no longer valid, and the cloud delay is clipped to a finite overflow penalty $T_{\rm cloud}^{\max}$, which is set to the profiled latency bound $L_{\max} = 120$ s, while excess packets are dropped. It is important to emphasize that the computational latency for executing the global PSKY dominance checks at the cloud server is assumed to be negligible and is absorbed into this minor constant baseline. This assumption is mathematically justified by the stark asymmetry in computational capabilities between the resource-constrained mobile edge nodes and the virtually infinite elasticity of the cloud server. Given that the wide-area wireless communication link introduces a significant latency bottleneck, the millisecond-level aggregation time at the cloud is orders of magnitude smaller. To maintain fundamental physical stability, the traffic intensity must theoretically satisfy $\rho_t = \Lambda_t(\boldsymbol{\alpha}_t)/\mu < 1$. In practical implementations, if $\rho_t \ge 1$, the system experiences queue overflow, resulting in explicit packet drops at the server, which serves as a critical boundary condition for our optimization agent.

\subsection{Data Utility and Approximate Probabilistic Skyline}

SA-PSKY operates in an exact-preserving regime when the adaptive local threshold does not exceed the baseline query threshold $\alpha_{\mathrm{query}}$. However, rigidly adhering to this constraint in resource-constrained mobile environments often induces severe network congestion or unacceptable computational latency. 

To ameliorate this systemic bottleneck, when $\alpha_{i,t} > \alpha_{\mathrm{query}}$, exact recall is no longer guaranteed, and the problem is treated as approximate probabilistic skyline analytics. In this controlled approximate regime, the decision agent is authorized to dynamically elevate the local filtering threshold $\alpha_{i,t}$ above $\alpha_{\mathrm{query}}$, enabling aggressive candidate set reduction during periods of high resource contention. This proactive data reduction fundamentally mitigates the risks of transmission queuing and computational overload.

To formalize the optimization, we explicitly quantify the positive informational gain of transmitting valid data, the filtering precision to prevent bandwidth waste, and the negative utility penalty associated with dropping critical candidates. We define the normalized data utility $\tilde{\mathcal{U}}_{t} \in [0,1]$, hereafter referred to as the hit rate, as the ratio of true global skyline candidates correctly identified and transmitted. We introduce filtering precision $\mathcal{P}_t \in [0,1]$ to measure the density of valid skylines within the transmitted packets. Conversely, the precision loss is quantified by the false negative rate $\bar{\epsilon}_{t}$, representing the ratio of true global skyline objects erroneously pruned during local filtering.

\subsection{Multi-Objective Optimization Problem Formulation}
The objective of the SA-PSKY framework is to derive a learned filtering policy that maximizes net system utility while strictly adhering to physical network capacity limits. We formalize this as a discrete-time optimization problem.

\subsubsection{System Objective Function}
We define the global objective function $J$ as the expected sum of instantaneous physical utility across a finite sequence of discrete time slots $T_{\max}$:
\begin{equation}\label{eq:objective_function}
    \max_{\boldsymbol{\pi}} J = \mathbb{E} \left[ \sum_{t=1}^{T_{\max}} U_{\rm sys}(t, \boldsymbol{\alpha}_t) \right],
\end{equation}
where $U_{\rm sys}(t, \boldsymbol{\alpha}_t)$ denotes the net system utility obtained by applying the joint action $\boldsymbol{\alpha}_t$ at time $t$.

\subsubsection{Decomposition of Instantaneous Utility}
The net system utility $U_{\rm sys}$ reflects the fundamental physical trade-off between the informational value of transmitted data and the consumed computational and temporal resources. It is formulated as the raw data utility minus the weighted physical costs:
\begin{equation}\label{eq:instantaneous_utility}
    U_{\rm sys}(t) = \sum_{i=1}^K \mathcal{U}_{i,t} - \left( \eta_1 \sum_{i=1}^K T_{\rm comp}(i, \alpha_{i,t}) + \eta_2 L_{\rm sys}(t) \right).
\end{equation}

Here, $\sum \mathcal{U}_{i,t}$ represents the absolute count of true global skyline candidates correctly identified and transmitted. The parameters $\eta_1$ and $\eta_2$ are application-specific conversion factors mapping local computation time and global system latency into a unified utility metric. 

Crucially, the absolute system latency $L_{\rm sys}(t)$ is strictly formulated as:
\begin{equation}\label{eq:system_latency}
    L_{\rm sys}(t) = \max_{i} \{ T_{\rm comp}(i, \alpha_{i,t}) \} + \sum_{i=1}^K T_{\rm trans}(i, \alpha_{i,t}) + T_{\rm cloud}(\rho_t).
\end{equation}
In this formulation, the $\max_{i} \{ \cdot \}$ operator captures the parallel execution bottleneck at the edge, where total processing latency is dominated by the slowest node due to the synchronous nature of the candidate aggregation process. The term $\sum_{i=1}^K T_{\rm trans}(i, \alpha_{i,t})$ accounts for the cumulative serialization delay encountered when multiple edge nodes contend for the bandwidth-constrained uplink channel. Finally, the $T_{\rm cloud}(\rho_t)$ term accounts for the processing and waiting time at the broker, dynamically driven by the aggregate traffic intensity $\rho_t$ derived from the Poisson arrival process.

\subsubsection{Constraint Set and Physical Drop Mechanism}
To enforce operational stability, the optimization is evaluated with the following continuous action bounds and network capacity limits:
\begin{equation}
    \text{with} \quad 
    \begin{cases}
        \alpha_{i,t} \in \left[0, 1\right], \quad \forall i \\
        \rho_t = \dfrac{\sum_{i=1}^K \lambda_{i,t} \cdot \sigma_i(\alpha_{i,t})}{\mu} < 1
    \end{cases}
\end{equation}
as the nominal stability condition; violations are modeled as overflow events and penalized accordingly. The physical network environment mandates a hard packet drop policy. If the joint action leads to a queue overflow state ($\rho_t \ge 1$), the broker discards overflowing packets.

\section{The Proposed Self-Adaptive Probabilistic Skyline Analytics Framework}
\label{sec:proposed_framework}

The proposed SA-PSKY framework introduces an autonomous mechanism to navigate the multi-dimensional optimization space of distributed data filtering. By modeling the orchestration between heterogeneous edge nodes and the cloud broker as a MDP, the system achieves an adaptive balance between local computational expenditure and global network latency, effectively neutralizing the performance bottlenecks inherent in distributed query processing. 

A critical architectural design choice in SA-PSKY is the deployment of a centralized DRL agent at the cloud broker. The cloud broker is used only for control-plane orchestration, whereas query data processing and local filtering remain at the edge. This centralized agent synthesizes the global system state $s_t$ to derive an effective joint action vector $\boldsymbol{\alpha}_t$, providing a unified control policy for all edge nodes. The communication overhead associated with policy synchronization is assumed to be negligible relative to the massive data-plane traffic, as control parameters are seamlessly encapsulated within existing protocol control packets.

Furthermore, unlike traditional optimization paradigms such as dynamic programming or model predictive control, which necessitate precise environmental transition matrices, the proposed approach is strictly model free. This characteristic is essential, as the extreme spatio-temporal heterogeneity, \textit{Non-Independent and Identically Distributed} (Non-IID) data characteristics, and stochastic shifts in IoE streams render the explicit modeling of the global state transition dynamics computationally intractable.

\subsection{MDP Formulation for Threshold Control}

We formalize the interaction between the edge-assisted query operator and the dynamic sensing environment as a tuple $(\mathcal{S}, \mathcal{A}, \mathcal{R}, \gamma)$.

\begin{itemize}
    \item \textbf{State Space} ($\mathcal{S}$): The state vector $s_t$ at time $t$ provides a comprehensive representation of the operational status, synthesized from distributed telemetry:
    \begin{equation}
        s_t = \{Q_t, \Omega_{t-1}, \rho_{t-1}, \mathcal{D}_t, \boldsymbol{\lambda}_t\},
    \end{equation}
    where $Q_t$ denotes the instantaneous broker queue length, $\mathcal{D}_t$ is the spatial distribution feature vector, and $\boldsymbol{\lambda}_t$ denotes the normalized data arrival rates from edge nodes. Crucially, $\Omega_{t-1}$ and $\rho_{t-1}$ represent the observed global data selectivity and system congestion pressure from the previous control interval, ensuring strict causal consistency in the Markov decision process.

    To capture distribution dynamics without the computational overhead of conventional \textit{Spatial Skyline Queries} (SSQ), $\mathcal{D}_t$ is formulated as a lightweight multi-dimensional representation, $\mathcal{D}_t = \{\mathbf{G}_t, \mathbf{V}_t\}$. Adopting spatial reference techniques adapted for the \textit{Internet of Vehicles} (IoV) \cite{Graham2023Edge}, we construct a virtual center as a heuristic proxy for spatial data concentration. First, let $\mathbf{c}_{i,t} \in \mathbb{R}^d$ denote the expected local feature centroid of active objects within $\mathcal{W}_i$:
    \begin{equation}
        \mathbf{c}_{i,t} = \frac{1}{|\mathcal{W}_i|} \sum_{u \in \mathcal{W}_i} \sum_{j=1}^{m} P(u_j) \cdot u_j.
    \end{equation}
    The global virtual center $\mathbf{g}_t \in \mathbb{R}^d$ is derived as the geometric mean of all local centroids, $\mathbf{g}_t = \frac{1}{K} \sum_{i=1}^{K} \mathbf{c}_{i,t}$. We characterize spatial crowding pressure using the centroid distance profile $\mathbf{G}_t = [G_{1,t}, \dots, G_{K,t}]^T \in \mathbb{R}^K$, where $G_{i,t} = \|\mathbf{c}_{i,t} - \mathbf{g}_t\|_2$.

    To address spatio-temporal shifts, we extract Pearson correlation coefficients to form the correlation metric vector $\mathbf{V}_t = [V_{1,t}, \dots, V_{K,t}]^T \in \mathbb{R}^K$, where $V_{i,t}$ quantifies the correlation between the historical trajectory of the local centroid and the global center over a short-term observation horizon $\Delta T=10$:
    \begin{equation}
        V_{i,t} = \frac{\sum_{\tau=t-\Delta T}^{t} (\mathbf{c}_{i,\tau} - \bar{\mathbf{c}}_i) \cdot (\mathbf{g}_\tau - \bar{\mathbf{g}})}{\sqrt{\sum_{\tau=t-\Delta T}^{t} \|\mathbf{c}_{i,\tau} - \bar{\mathbf{c}}_i\|_2^2} \cdot \sqrt{\sum_{\tau=t-\Delta T}^{t} \|\mathbf{g}_\tau - \bar{\mathbf{g}}\|_2^2}},
    \end{equation}
    where $\bar{\mathbf{c}}_i$ and $\bar{\mathbf{g}}$ signify their respective temporal expectations. By aggregating these profiles, the spatial feature space is condensed into $\mathcal{D}_t \in \mathbb{R}^{2K}$. These sub-sampling operations target lightweight feature extraction \cite{10308406} under a bounded complexity of $O(N \cdot d + K \cdot \Delta T)$, ensuring the feature extraction process remains highly scalable for large-scale, dynamic network topologies.

    \item \textbf{Action Space} ($\mathcal{A}$): The action $a_t$ represents the adjustment of the threshold vector $\boldsymbol{\alpha}_{t} = \{\alpha_{1,t}, \dots, \alpha_{K,t}\}$. SA-PSKY employs a continuous action space where each $\alpha_{i,t} \in [0, 1]$, enabling infinitesimal, fine-grained adjustments to the pruning probability intensity.

    \item \textbf{Reward Function} ($\mathcal{R}$): The reward $r_t$ is designed to reflect the multi-objective utility encompassing computation, latency, and query accuracy. To resolve scale discrepancies among these heterogeneous physical metrics and facilitate convergence stability, $r_t$ is computed via a normalized reward shaping mechanism, further detailed in the subsequent section.
\end{itemize}

\subsection{Deep Deterministic Policy Gradient Architecture}

The continuous nature of the threshold adjustment problem necessitates an Actor-Critic architecture to facilitate smooth action mapping within the high-dimensional threshold space. The SA-PSKY agent employs four distinct neural networks: the actor $\mu(s|\theta^\mu)$, the critic $Q(s, a|\theta^Q)$, and their respective target networks $\mu'$ and $Q'$.

\begin{itemize}
    \item \textbf{Actor Network}: This network parameterizes the control policy by mapping state vectors to deterministic threshold actions. It comprises an input layer, three hidden layers with ReLU activation, and a Sigmoid output layer featuring $K$ neurons, ensuring the joint action vector $\boldsymbol{\alpha}_t$ strictly resides in the unit hypercube $\alpha_{i,t} \in [0, 1]$ for all edge nodes.
    \item \textbf{Critic Network}: The critic approximates the action-value function $Q(s, a)$. To effectively model the non-linear coupling between operational states and threshold actions, the action vector $\boldsymbol{\alpha}_t$ is integrated with state features at the second hidden layer.
\end{itemize}

\subsection{MDP Mapping and State-Aware Reward Shaping}

While \eqref{eq:objective_function} formalizes the physical optimization objective, direct integration of raw system metrics into the MDP poses significant training challenges. To prevent variables with disparate magnitudes from destabilizing gradient updates and to resolve the degenerate zero-selectivity trap, we implement a utility-driven, normalized reward shaping structure augmented with the \textit{State-Aware Adaptive Weighting} (SAAW) mechanism. 

The agent optimizes the shaped instantaneous reward $r_t$, which serves as a normalized reward-shaping surrogate of the physical objective in \eqref{eq:objective_function}--\eqref{eq:instantaneous_utility}. This surrogate is designed to preserve the relative trade-offs while improving numerical conditioning and boundary awareness during RL training:
\begin{align}\label{eq:reward}
    r_{t} &= \tilde{\mathcal{U}}_{t} + \zeta \mathcal{P}_t - \bigg( w_1 \cdot \frac{\sum_{i=1}^K T_{\rm comp}(i, \alpha_{i,t})}{C_{\max}} + w_2(\rho_t) \cdot \frac{L_{\rm sys}(t)}{L_{\max}} \notag\\
    &\quad + w_3(\rho_t) \cdot \sqrt{\bar{\epsilon}_{t}} + w_4(\rho_t) \cdot (\Omega_{t} + \Omega_{t}^2) + \mathcal{P}_{\exp} \bigg),
\end{align}
where $\rho_t$ denotes the post-action traffic pressure, and $\mathcal{P}_{\exp} = \lambda_{\exp} \cdot \max(0, \exp(\rho_t - 1) - 1)$ denotes the exponential pressure penalty, activated exclusively upon queue overflow ($\rho_t \ge 1$), with $\lambda_{\exp}$ serving as a scaling coefficient to modulate penalty intensity.  Furthermore, $\tilde{\mathcal{U}}_{t}$ and $\mathcal{P}_t$ denote normalized data utility and filtering precision, weighted by scalar $\zeta$ to balance their relative influence. Normalization constants $C_{\rm max}$ and $L_{\rm max}$ are derived from empirical worst-case profiling to ensure training stability. A square root transformation $\sqrt{\bar{\epsilon}_{t}}$ is applied to the false negative penalty to amplify the cost of critical information loss, thereby enforcing high query fidelity. Concurrently, the compound linear-quadratic term $(\Omega_{t} + \Omega_{t}^2)$ imposes an escalating penalty on bandwidth overconsumption, effectively discouraging uplink channel saturation.

To ensure resilience against traffic surges, the weighting coefficients are dynamically driven by the real-time queue pressure $\rho_t = \min(\Lambda(\alpha_t)/\mu, 1.0)$. We define the SAAW configuration as follows:
\begin{itemize}
    \item \textbf{Computation Weight ($w_1$)}: Constant at $0.2$, maintaining a stable baseline constraint on local edge CPU resources.
    \item \textbf{Latency Weight ($w_2$)}: $0.1 + 0.2\rho_t$, penalizing end-to-end delay proportionally to the network congestion level.
    \item \textbf{Accuracy Weight ($w_3$)}: $0.6 - 0.4\rho_t$, prioritizing precision during idle network states, while adaptively trading off accuracy to preserve stability during peak loads.
    \item \textbf{Bandwidth Weight ($w_4$)}: $0.1 + 0.2\rho_t$, imposing heavy penalties on redundant transmissions as the uplink channel approaches saturation.
\end{itemize}

These coefficients define the adopted state-aware reward envelope and are calibrated to satisfy the simplex normalization $\sum_{i=1}^4 w_i = 1.0$ across the entire operational spectrum of queue pressure $\rho_t \in [0, 1.0]$, ensuring normalized weighting within the optimization bounds:
\begin{equation}
    \sum_{i=1}^4 w_i = 0.2 + (0.1 + 0.2\rho_t) + (0.6 - 0.4\rho_t) + (0.1 + 0.2\rho_t) = 1.0.
\end{equation}

The specific bounds and intercepts of these coefficients are calibrated based on empirical profiling the extreme operational boundaries of the mobile edge system. It is crucial to emphasize that these baseline parameters serve merely as a foundational envelope for the reward structure, rather than rigid, scenario-specific constants. The core intelligence of the SA-PSKY framework does not rely on a perfectly tuned set of static weights. Instead, it leverages the state-aware mechanism to continuously and autonomously recalibrate these coefficients in response to the real-time traffic intensity $\rho_t$. 

Within this adaptive envelope, the computation weight $w_1$ acts as a constant anchor representing the physical processing limit of edge devices. The accuracy weight $w_3$ dynamically spans $[0.2, 0.6]$, ensuring that query fidelity dominates the optimization objective during low-traffic periods ($\rho_t \to 0$). Conversely, as the broker approaches saturation ($\rho_t \to 1.0$), the SAAW mechanism smoothly throttles the accuracy priority down to 0.2 while autonomously amplifying the latency and bandwidth penalties ($w_2, w_4 \to 0.3$). This specific dynamic formulation structurally compels the agent to elevate $\alpha_{i,t}$ and execute proactive data reduction under varying mobility workloads, thereby preventing severe network boundary violations without falling into the trivial zero-selectivity trap.

\subsection{Policy Optimization and Stability Mechanisms}

The SA-PSKY agent utilizes the Actor-Critic architecture to iteratively refine its threshold control policy. The Critic network, parameterized by $\theta^Q$, approximates the optimal action-value function by minimizing the \textit{Mean Squared Bellman Error} (MSBE) over a mini-batch of $M$ transitions:
\begin{equation}
    L(\theta^Q) = \frac{1}{M} \sum_{j=1}^{M} (y_j - Q(s_j, a_j|\theta^Q))^2,
\end{equation}
where $y_j = r_j + \gamma Q'(s_{j+1}, \mu'(s_{j+1}|\theta^{\mu'})|\theta^{Q'})$ denotes the target value for the $j$-th sample. Simultaneously, the Actor network, parameterized by $\theta^\mu$, is updated to maximize the expected cumulative return via the sampled policy gradient:
\begin{equation}
    \nabla_{\theta^\mu} J \approx \frac{1}{M} \sum_{j=1}^{M} \nabla_a Q(s, a|\theta^Q)|_{s=s_j, a=\mu(s_j)} \nabla_{\theta^\mu} \mu(s|\theta^\mu)|_{s_j}.
\end{equation}

To ensure training stability and rapid convergence within non-stationary IoE environments, the framework incorporates two pivotal mechanisms:

\begin{itemize}
    \item \textbf{Prioritized Experience Replay (PER)}~\cite{per2016}: Beyond uniform sampling, we implement a PER buffer $\mathcal{B}$ where transitions are assigned priority according to their \textit{Temporal-Difference} (TD) error. This configuration empowers the agent to prioritize the learning of high-impact system states, particularly during abrupt network congestion or severe distribution shifts.
    \item \textbf{Soft Target Updates}: To suppress policy oscillation originating from volatile target value estimation, the target networks $\theta^{\mu'}$ and $\theta^{Q'}$ are updated using an incremental tracking mechanism:
    \begin{equation}
        \begin{cases}
            \theta^{Q'} \leftarrow \tau \theta^Q + (1-\tau) \theta^{Q'} \\
            \theta^{\mu'} \leftarrow \tau \theta^\mu + (1-\tau) \theta^{\mu'}
        \end{cases}
    \end{equation}
    where $\tau \ll 1$ denotes the target smoothing coefficient. This mechanism ensures that target values evolve gradually, maintaining decision consistency in time-sensitive, resource-constrained edge environments.
\end{itemize}

\subsection{Exploration via Ornstein-Uhlenbeck Process}

Within the threshold control domain, simple additive white noise often proves insufficient due to the inherent temporal correlations in sensor data streams. We employ the \textit{Ornstein-Uhlenbeck} (OU) process $\mathcal{N}_t$ to inject temporally correlated exploration noise into the action selection phase:
\begin{equation}
    a_t = \text{clip}(\mu(s_t|\theta^\mu) + \mathcal{N}_t, \alpha_{\min}, \alpha_{\max}).
\end{equation}
The OU process facilitates mean-reverting exploration, ensuring that the agent investigates the threshold space smoothly within the strictly bounded domain $[\alpha_{\min}, \alpha_{\max}] = [0, 1]$. This stochastic exploration strategy allows the system to accurately map the long-term impact of threshold fluctuations on broker queue dynamics and overall system utility.

\subsection{Algorithmic Implementation and Training Procedure}

The implementation adopts an off-policy training regime to ensure superior sample efficiency and environmental adaptability. The complete training procedure is detailed in Algorithm~\ref{alg:sa_psky}, which orchestrates the interaction between the centralized DDPG agent and the dynamic mobile sensing environment. The procedure is structured into three pivotal phases.

\subsubsection{Phase 1: Initialization}
The process commences by initializing the primary Actor network $\mu(s|\theta^{\mu})$ and Critic network $Q(s, a|\theta^{Q})$ with random weights. To ensure stable convergence, we instantiate target networks, denoted as $\theta^{\mu'}$ and $\theta^{Q'}$, which serve as delayed references for value estimation. Concurrently, a PER buffer $\mathcal{B}$ is initialized to store transition tuples, facilitating non-uniform sampling of high-impact historical experiences.

\subsubsection{Phase 2: Interaction and Exploration}
At each time step $t$, the agent observes the current system state $s_t$. To traverse the continuous action space effectively, we inject temporally correlated noise $\mathcal{N}_t$ (via the OU process) into the deterministic action generated by the Actor. The resultant action $a_t$ is clipped to the valid operational range $[\alpha_{\min}, \alpha_{\max}]$ and applied as the filtering threshold vector $\boldsymbol{\alpha}_t$. The system subsequently transitions to state $s_{t+1}$ and returns a scalar reward $r_t$. This transition tuple $(s_t, a_t, r_t, s_{t+1})$ is stored in $\mathcal{B}$ with maximal priority, ensuring its inclusion in the initial optimization iterations.

\subsubsection{Phase 3: Network Optimization}
Once the buffer $\mathcal{B}$ accumulates a sufficient volume of samples, the optimization loop proceeds:
\begin{itemize}
    \item \textbf{Critic Update}: A mini-batch of $M$ transitions is sampled based on assigned priorities. We compute importance sampling weights $\iota_j$ to mitigate estimation bias. The Critic network minimizes the weighted \textit{Mean Squared Bellman Error} (MSBE) between the predicted Q-value and the target $y_j$, where $y_j$ is calculated via the Bellman equation using the target networks.
    \item \textbf{Actor Update}: The Actor is updated utilizing the \textit{Deterministic Policy Gradient} (DPG) method. The gradient $\nabla_{\theta^{\mu}} J$ is derived by applying the chain rule to the Critic output with respect to the action, thereby adjusting policy parameters to maximize the Q-value.
    \item \textbf{PER Priority Update}: Priorities within $\mathcal{B}$ are updated according to the absolute \textit{Temporal-Difference} (TD) error $|y_j - Q(s_j, a_j|\theta^{Q})|$, ensuring that experiences representing sharp environmental transitions are preferentially replayed.
    \item \textbf{Soft Target Update}: The weights of the target networks are incrementally tracked toward the primary networks using a smoothing coefficient $\tau$. This mechanism suppresses value oscillation and ensures decision consistency across time-sensitive edge-cloud operations.
\end{itemize}

\begin{algorithm2e}[t]
\small
\caption{SA-PSKY Threshold Optimization via DDPG}
\label{alg:sa_psky}
\SetAlgoLined
\SetKwInOut{Input}{Input}
\SetKwInOut{Output}{Output}

\Input{Initial state $s_1$, Episodes $E_{\max}$, Steps $T_{\max}$, Learning rates $\eta_{\mu}, \eta_{Q}$, Soft update $\tau$, Discount factor $\gamma$}
\Output{Optimized Policy Parameters $\theta^{\mu}$, Critic Parameters $\theta^{Q}$}

\BlankLine
\tcp{Phase 1: Initialization}
Initialize Actor $\mu(s|\theta^{\mu})$ and Critic $Q(s, a|\theta^{Q})$ with random weights\;
Initialize target networks $\theta^{\mu'} \leftarrow \theta^{\mu}$ and $\theta^{Q'} \leftarrow \theta^{Q}$\;
Initialize PER buffer $\mathcal{B}$ and exploration noise $\mathcal{N}$\;

\BlankLine
\For{episode = 1 \KwTo $E_{\max}$}{
    Reset mobile environment and receive initial state $s_1$\;
    \For{$t = 1$ \KwTo $T_{\max}$}{
        \tcp{Phase 2: Interaction}
        Select action $a_t = \text{clip}(\mu(s_t|\theta^{\mu}) + \mathcal{N}_t, \alpha_{\min}, \alpha_{\max})$\;
        Execute action $a_t$ (set local thresholds $\boldsymbol{\alpha}_t = a_t$)\;
        Evaluate post-action metrics, compute reward $r_t$, and observe state $s_{t+1}$\;
        Store transition $(s_t, a_t, r_t, s_{t+1})$ in $\mathcal{B}$ with maximal priority\;
        
        \If{$\text{size of } \mathcal{B} > \text{Batch Size } M$}{
            \tcp{Phase 3: Optimization}
            Sample mini-batch $M$ from $\mathcal{B}$ indexed by $j$ with weights $\iota_j$\;
            Set target $y_j = r_j + \gamma Q'(s_{j+1}, \mu'(s_{j+1}|\theta^{\mu'})|\theta^{Q'})$\;
            
            \BlankLine
            \tcp{Update Critic Network}
            Update Critic by minimizing weighted MSBE: \\
            $L = \frac{1}{M} \sum_{j=1}^{M} \iota_j (y_j - Q(s_j, a_j|\theta^{Q}))^2$\;
            
            \BlankLine
            \tcp{Update Actor Network}
            $\nabla_{\theta^{\mu}} J \approx \frac{1}{M} \sum_{j=1}^{M} \nabla_a Q(s, a|\theta^{Q})|_{s=s_j, a=\mu(s_j)} \nabla_{\theta^{\mu}} \mu(s|\theta^{\mu})|_{s_j}$\;

            \BlankLine
            \tcp{Update PER Priorities}
            Update priorities in $\mathcal{B}$ via $|y_j - Q(s_j, a_j|\theta^{Q})|$\;
            
            \BlankLine
            \tcp{Soft Update Target Networks}
            $\theta^{\mu'} \leftarrow \tau \theta^{\mu} + (1 - \tau) \theta^{\mu'}$\;
            $\theta^{Q'} \leftarrow \tau \theta^{Q} + (1 - \tau) \theta^{Q'}$\;
        }
        $s_t \leftarrow s_{t+1}$\;
    }
    \tcp{Decay exploration noise}
    Update $\mathcal{N}$ for the next episode\;
}
\end{algorithm2e}

\subsection{Computational Complexity and Pruning Efficiency}

The computational overhead at each edge node is predominantly determined by the complexity of probabilistic dominance comparisons. For a local active dataset of size $N$ (where $N \le W_{\max}$) with $m$ discrete instances distributed across $d$ dimensions, the brute-force processing complexity is strictly bounded by $O(N^2 \cdot m^2 \cdot d)$. This severe non-linear complexity necessitates an adaptive pruning mechanism to prevent CPU saturation in resource-constrained edge environments.

By enforcing the dynamic filtering threshold $\alpha$, the SA-PSKY framework curtails the redundant search space. Let $N_{\alpha}$ denote the number of candidate objects whose local skyline probability satisfies the strict threshold condition ($N_{\alpha} \ll N$). The pruning efficiency factor $\Phi(\alpha)$, which quantifies the candidate retention ratio, can be theoretically approximated as $\Phi(\alpha) \approx N_{\alpha} / N$. Consequently, the average-case computational complexity is effectively reduced to $O(N \cdot N_{\alpha} \cdot m^2 \cdot d)$. This theoretical analysis motivates the monotonic relationship between filtering selectivity and computational workload, while the empirical formulation in \eqref{eq:comp_surrogate} provides a tractable surrogate used in the control optimization.

Crucially, this theoretical reduction establishes $\alpha$ as the primary regulator of local computational expenditure. As the underlying spatial data distribution shifts, the baseline candidate survival count $N_{\alpha}$ tends to inflate. More critically, as feature dimensionality $d$ expands, the probability of mutual non-dominance is inherently exacerbated due to the curse of dimensionality, which further elevates $N_{\alpha}$. This structural challenge compels the DRL agent to adjust $\alpha$ dynamically, underscoring the necessity of the multi-objective reward function in orchestrating the critical balance between mitigating the dual physical burdens (local computation and global transmission) and preserving query fidelity.

\subsection{Theoretical Rationale for Training Stability}

We do not claim theoretical convergence of the off-policy actor-critic training process. Rather, PER, importance sampling weights, and soft target updates are employed as practical stabilization mechanisms. These components work synergistically to provide a smoother gradient flow and prevent severe policy oscillations near strict penalty boundaries. The effectiveness of this stability-oriented design will be comprehensively evaluated and demonstrated empirically in Section~\ref{sec:experiments}.

\section{Experimental Evaluation}
\label{sec:experiments}

In this section, we conduct a comprehensive evaluation of the proposed SA-PSKY framework. A critical aspect of this study is the analytical assessment of the trade-off between the computationally intensive dominance checks required for uncertain data and the constrained network bandwidth inherent in IoE environments.

\subsection{Simulation Environment and Settings}
To simulate a realistic mobile edge computing environment, network communication between edge nodes and the cloud broker is modeled upon the enhanced \textit{Machine Type Communication} (eMTC) architecture. The uplink bandwidth is strictly constrained to 1 Mbps, faithfully emulating the congested wireless channels typical of massive mobile edge deployments. 

To reflect lightweight data collection practices, each uncertain object is encapsulated via the \textit{Constrained Application Protocol} (CoAP). The average packet size is set to 1 Kbit, representing a standard payload for compact JSON or CBOR-formatted sensor readings. Crucially, processing these uncertain objects necessitates probabilistic dominance checks with a time complexity of $O(N^2 \cdot m^2 \cdot d)$, where $m$ denotes the number of instances and $d$ represents the feature dimension. This computational overhead renders local processing latency a non-negligible factor in evaluating end-to-end performance.

One simulation step represents a fixed control interval of $\Delta t = 1.0$ second. Consequently, all data arrival and service rates specified per step are mathematically equivalent to absolute rates per physical second, ensuring strict dimensional consistency when evaluating the queuing dynamics and physical latency metrics. All fundamental system variables, alongside the SA-PSKY deep reinforcement learning hyperparameters, are summarized in Table~\ref{tab:combined_parameters}. Based on these configurations, the DRL agent employs an \textit{Ornstein-Uhlenbeck} process to inject temporally correlated noise for policy exploration. This mechanism is essential for navigating the continuous action space of the proposed framework, ensuring a smooth and stable balance between exploration and exploitation. Furthermore, the normalization parameters $L_{\rm max}$ and $C_{\rm max}$, as detailed in the table, are established based on the worst-case empirical profiling of the centralized \textit{No-Filtering} baseline. These bounds guarantee that the shaped reward $r_t$ remains consistently scaled, thereby preventing gradient instability during policy updates.

To rigorously capture the spatial and temporal complexities inherent in practical mobile deployments, the dataset is engineered to exhibit extreme spatio-temporal heterogeneity and dynamic distribution shifts. This critical data imbalance is modeled across two distinct dimensions:
\begin{itemize}
    \item \textbf{Traffic Intensity Skew}: Rather than assuming a uniform generation rate, the local arrival rate $\lambda_i$ for each edge node is independently sampled from a continuous uniform distribution $\mathcal{U}(4.0, 12.0)$. This induces an asymmetric network load, forcing the central broker to manage highly unpredictable queuing dynamics.
    \item \textbf{Spatial Feature Skew}: Driven by the geographical separation of sensors, the multi-dimensional features observed by different edge nodes possess distinct statistical moments. Consequently, a static global threshold cannot simultaneously optimize the precision-bandwidth trade-off across all nodes, thereby establishing the explicit necessity for a state-aware dynamic thresholding mechanism.
\end{itemize}

\begin{table}[!t]
\centering
\caption{System Simulation and SA-PSKY DRL Parameters}
    \label{tab:combined_parameters}
    \setlength{\tabcolsep}{3pt}
    \begin{tabular}{ll}
        \hline
        \textbf{Parameter} & \textbf{Default Value} \\
        \hline
        \multicolumn{2}{c}{\textit{Neural Network Architecture}} \\ \hline
        Hidden Layers (Actor \& Critic) & 3 (400, 300, 200 neurons) \\
        Activation (Hidden / Output) & ReLU / Sigmoid \\
        Actor Learning Rate ($\eta_{\mu}$) & $10^{-4}$ \\
        Critic Learning Rate ($\eta_{Q}$) & $10^{-3}$ \\
        Mini-batch Size ($M$) & 128 \\ \hline
        \multicolumn{2}{c}{\textit{Optimization \& Exploration Dynamics}} \\ \hline
        Discount Factor ($\gamma$) & 0.99 \\
        Soft Target Update Rate ($\tau$) & 0.005 \\
        OU Noise Base Volatility ($\sigma$) & 0.20 \\
        OU Noise Minimum ($\sigma_{\min}$) & 0.05 \\
        OU Noise Decay Rate ($\gamma_{\sigma}$) & 0.996 \\ \hline
        \multicolumn{2}{c}{\textit{Prioritized Experience Replay (PER)}} \\ \hline
        Replay Buffer Capacity ($|\mathcal{B}|$) & $10^6$ transitions \\
        Priority Exponent ($\nu$) & 0.4 \\
        Initial IS Weight ($\beta_0$) & 0.4 (Annealed to 1.0) \\
        Small Constant ($\epsilon$) & $0.01$ \\ \hline
        \multicolumn{2}{c}{\textit{Network \& Data Constraints}} \\ \hline
        Total Data Volume & 50,000 objects \\
        Data Distribution & Spatio-temporally Heterogeneous \\
        Number of Edge Servers ($K$) & 5 \\
        Instances per Object ($m$) & 10 \\
        Data Dimensions ($d$) & 2 \\
        Average Object Size ($\omega$) & 1 Kbit \\
        Communication Overhead Factor ($\xi$) & 12 \\
        Network Bandwidth ($B$) & 1 Mbps \\
        Sliding Window Capacity ($W_{\max}$) & 500 \\ \hline
        \multicolumn{2}{c}{\textit{System Queuing \& Environment Dynamics}} \\ \hline
        Control Interval ($\Delta t$) & 1.0 s (1 simulation step) \\
        Local Arrival Rate ($\lambda_{i,t}$) & $\mathcal{U}(4.0, 12.0)$ objs/step \\
        Cloud Service Rate ($\mu$) & 50.0 objs/step \\
        Unit Computation Cost ($c_{\rm comp}$) & 0.005 s/obj \\
        Unit Transmission Cost ($c_{\rm trans}$) & 0.012 s \\
        Observation Horizon ($\Delta T$) & 10 steps \\
        Episode Length ($T_{\max}$) & 1000 steps \\
        Training Episodes ($E_{\max}$) & 1000 \\ 
        Utility Conversion Factors ($\eta_1, \eta_2$) & 1.0, 1.0 \\
        Precision Weighting Scalar ($\zeta$) & 0.5 \\ 
        Exponential Penalty Scalar ($\lambda_{\exp}$) & 1.0 \\ \hline
        \multicolumn{2}{c}{\textit{Normalization Constants Derived from Profiling}} \\ \hline
        $L_{\rm max}$ (Max End-to-End Latency) & 120.0 s \\
        $C_{\rm max}$ (Max Aggregate Cost) & 326.55 units \\ \hline
    \end{tabular}
\end{table}

To systematically evaluate policy convergence and resource efficiency, the performance of the proposed SA-PSKY framework is benchmarked against the following diverse baseline strategies:
\begin{enumerate}
    \item \textbf{No-Filtering (System Boundary)}: All uncertain objects are transmitted to the cloud broker without any local pruning. This scenario serves as the theoretical failure boundary to demonstrate the severe network congestion and queue overflow incurred when edge intelligence is absent.
    \item \textbf{Proportional-Integral-Derivative based Adaptive (PID-AT)}: A heuristic threshold controller. This baseline represents traditional non-learning control systems, exposing the limitations of reactive feedback loops when confronting non-stationary distribution shifts in data streams.
    \item \textbf{Deep Q-Network (DQN)}: A fundamental value-based DRL algorithm with a discretized action space, where the threshold $\alpha \in [0, 1]$ is quantized into fixed levels. This baseline highlights the performance bottlenecks introduced by quantization errors and isolates the specific benefits of continuous control.
    \item \textbf{Twin Delayed DDPG (TD3)}: An advanced actor-critic framework employing clipped double Q-learning. The inclusion of TD3 evaluates the impact of over-pessimistic value estimation, explicitly demonstrating the policy oscillations that occur near strict network penalty boundaries.
    \item \textbf{Proximal Policy Optimization (PPO)}: A state-of-the-art on-policy DRL algorithm. PPO is included to assess the behavior of an on-policy learner without experience replay under the same resource-constrained environment.
\end{enumerate}

\subsection{Performance Evaluation Metrics}
To comprehensively assess the effectiveness of the evaluated frameworks, we dissect system performance across query accuracy, resource optimization, and end-to-end latency dimensions.

\subsubsection{Accuracy and Resource Optimization Metrics}
Unlike the exact skyline approach which guarantees complete retrieval, the proposed framework introduces controlled approximation at the edge layer. Since the central broker performs an exhaustive global dominance check on all received candidates, no false positive objects are included in the final result set. Evaluation primarily focuses on the trade-off between retaining valid objects and minimizing data traffic. We quantify this using the following metrics:

\begin{itemize}
    \item \textbf{Total System Reward ($U_{\rm sys}$)}: The composite objective function score integrating data utility, computational latency penalties, and bandwidth congestion penalties, serving as the primary indicator of the Pareto-optimal balance achieved by the DRL agents.
    
    \item \textbf{Data Selectivity ($\Omega_t$)}: The ratio of objects passing local threshold filters to the total number of active objects across all edge sliding windows. Let $\mathcal{W}_i$ denote the active data window at node $i$, and $\mathcal{S}_i(\alpha_{i,t})$ denote the pruned candidate set. Global selectivity is defined as:
    \begin{equation}
        \Omega_t = \frac{\sum_{i=1}^K |\mathcal{S}_i(\alpha_{i,t})|}{\sum_{i=1}^K |\mathcal{W}_i|}. \notag
    \end{equation}
    This metric explicitly bounds the transmission overhead and directly drives the bandwidth penalty in the state-aware reward function.
    
    \item \textbf{Hit Rate (Empirical Recall, $\tilde{\mathcal{U}}_{t}$)}: The fraction of true global skyline objects successfully retained and transmitted by the adaptive local filters. Let $\mathcal{G}_{\rm exact}$ denote the absolute ground-truth set of global probabilistic skyline objects within the current time window. The empirical hit rate is formulated as:
    \begin{equation}
        \tilde{\mathcal{U}}_{t} = \frac{\left| \mathcal{G}_{\rm exact} \cap \left( \bigcup_{i=1}^K \mathcal{S}_i(\alpha_{i,t}) \right) \right|}{|\mathcal{G}_{\rm exact}|}. \notag
    \end{equation}
    This metric strictly quantifies the informational utility preserved by the agent, where a value approaching $1.0$ indicates zero false negatives during the aggressive local pruning phase.
\end{itemize}

\subsubsection{Latency Metrics}
The total system responsiveness is decomposed into its primary operational bottlenecks based on the formulations established in Section~\ref{sec:system_model}:

\begin{itemize}
    \item \textbf{Transmission Latency ($T_{\rm trans}$)}: The time required to transmit the filtered candidate sets from all edge nodes to the cloud broker over the shared bandwidth-constrained uplink channel.
    \item \textbf{Computation Latency ($T_{\rm comp}$)}: The execution time consumed by the instance-level probabilistic dominance checks at the edge layer, which is governed by the maximum processing time among parallel edge nodes.
    \item \textbf{End-to-End System Latency ($L_{\rm sys}(t)$)}: The aggregate system delay experienced at time step $t$, as formally defined in \eqref{eq:system_latency}. In our empirical evaluation under bandwidth-constrained eMTC configurations, the uplink bandwidth acts as a strict physical bottleneck, ensuring that the filtered candidate arrival rate generally remains below the broker service capacity ($\Lambda_t < \mu$). Consequently, the absolute queuing delay approaches zero, and the cloud processing time $T_{\rm cloud}(\rho_t)$ degrades into a negligible constant base service time. However, it is critical to note that the traffic intensity $\rho_t = \min(\Lambda_t/\mu, 1.0)$ acts as a vital state and constraint signal for the adaptive controller rather than a primary latency contributor. Since a constant delay offset does not alter the gradient direction in policy optimization, the effective system latency managed by the DRL agent is characterized by:
    \begin{equation}
        L_{\rm sys}(t) \approx \max_{i} \{ T_{\rm comp}(i, \alpha_{i,t}) \} + \sum_{i=1}^K T_{\rm trans}(i, \alpha_{i,t}). \notag
    \end{equation}
    It is important to note that $L_{\rm sys}(t)$ measures data-plane query latency, excluding the control-plane DRL inference overhead at the broker. This formulation explicitly captures the dual physical latency bottlenecks, parallel edge computation and serialized channel transmission, that the DRL agent must jointly minimize.
\end{itemize}

\subsection{Comparative Evaluation and Convergence Analysis}
\label{subsec:comparative_evaluation}

To evaluate the proposed SA-PSKY framework, we benchmark it against four baselines: TD3, PPO, DQN, and a heuristic PID-AT controller. The evaluation encompasses training convergence dynamics and multi-objective Pareto analysis presented in Fig.~\ref{fig:convergence}, alongside system delay profiling during inference in Fig.~\ref{fig:latency}.

\subsubsection{Multi-Objective Convergence and Policy Stability}

\begin{figure}[!t]
    \centering
    \subfloat[Total System Reward $U_{\rm sys}$]{
        \label{fig:convergence:a}
        \includegraphics[width=0.48\linewidth]{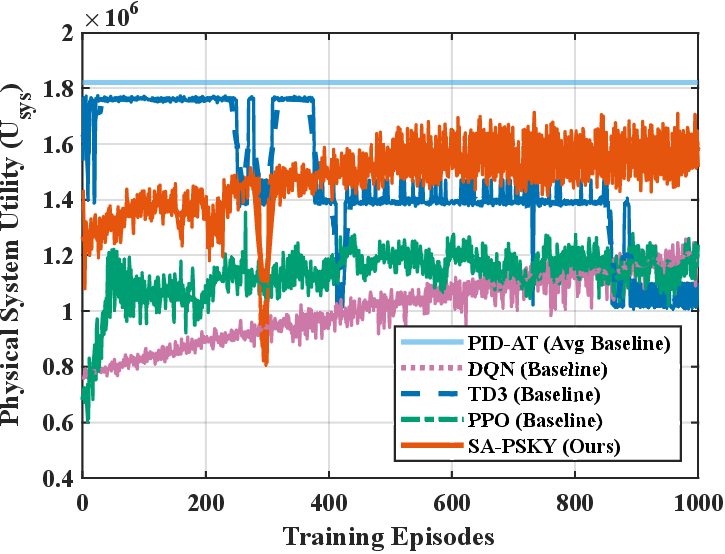}
    }%
    \subfloat[Data Selectivity $\Omega_t$]{
        \label{fig:convergence:b}
        \includegraphics[width=0.48\linewidth]{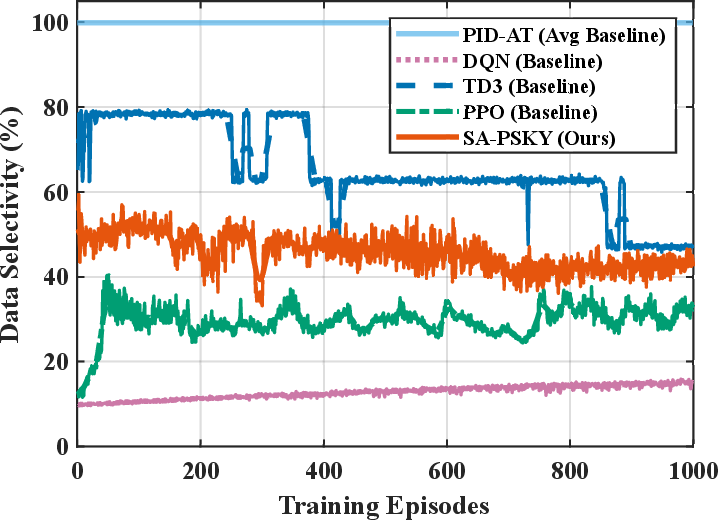}
    }\\
    \subfloat[Hit Rate $\tilde{\mathcal{U}}_{t}$]{
        \label{fig:convergence:c}
        \includegraphics[width=0.48\linewidth]{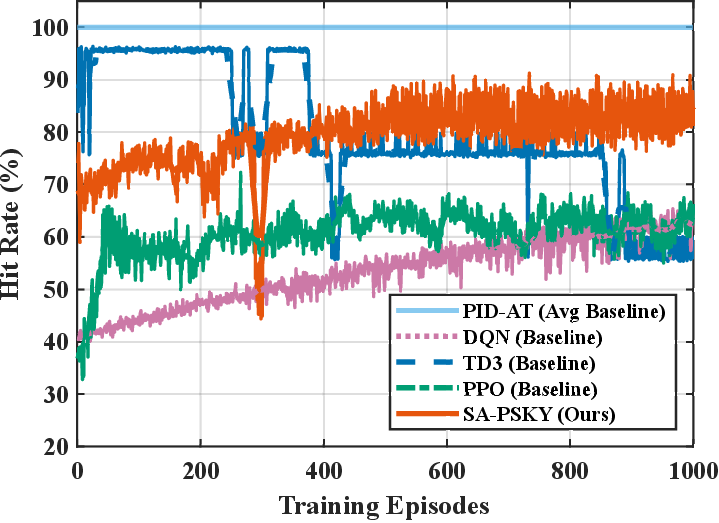}
    }%
    \subfloat[Pareto-Efficiency Analysis]{
        \label{fig:convergence:d}
        \includegraphics[width=0.47\linewidth]{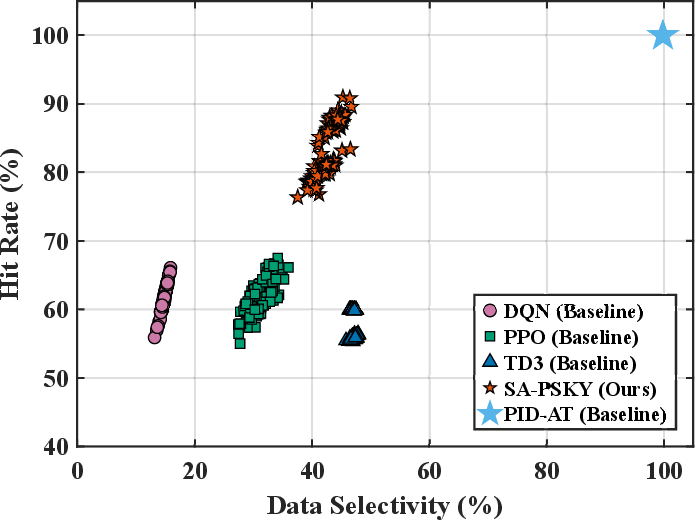}
    }
    \caption{Comprehensive evaluation of the proposed SA-PSKY against baseline algorithms. (a) Total system reward convergence. (b) Data selectivity dynamics. (c) Hit rate performance. (d) Pareto-efficiency analysis visualization.}
    \label{fig:convergence}
\end{figure}

As illustrated in Fig.~\ref{fig:convergence:a}, SA-PSKY exhibits a superior convergence trajectory, consistently achieving the highest total system reward among the evaluated reinforcement learning algorithms. This mechanism is further substantiated in Fig.~\ref{fig:convergence:b} and Fig.~\ref{fig:convergence:c}. Initially, SA-PSKY establishes a robust baseline to effectively avoid queue overflow penalties. As exploratory noise anneals, the agent learns to move toward a Pareto-efficient operating region, stabilizing data selectivity within the 40\% to 45\% range while maintaining a competitive hit rate approaching 85\%.

The objective space visualization in Fig.~\ref{fig:convergence:d} confirms this dominance. The PID-AT controller provides a high-fidelity empirical reference, achieving a 99.97\% hit rate at the excessive cost of 99.85\% data selectivity. In contrast, SA-PSKY actively converges toward a high-quality Pareto-efficient region. This distribution indicates a superior trade-off in the selectivity-hit-rate objective space, maintaining high query reliability while significantly optimizing data selectivity.

Conversely, TD3 suffers from severe policy oscillation and converges to a suboptimal operating regime. Failing to harmonize multi-objective constraints, its policy convergence stagnates at a suboptimal strategy with a hit rate below 60\%. Meanwhile, the stochastic nature of PPO and the coarse discrete granularity of DQN restrict them to overly conservative local optima. While these baselines remain stable, they limit selectivity to approximately 30\% and 15\% respectively, lacking the precision to navigate the continuous Pareto surface and yielding hit rates barely exceeding 60\%.

\subsubsection{The Cost of Intelligence: Delay Profiling and Adaptability}
\begin{figure*}[!t]
    \centering
    \subfloat[System Latency Components]{
        \label{fig:latency:a}
        \includegraphics[width=0.315\linewidth]{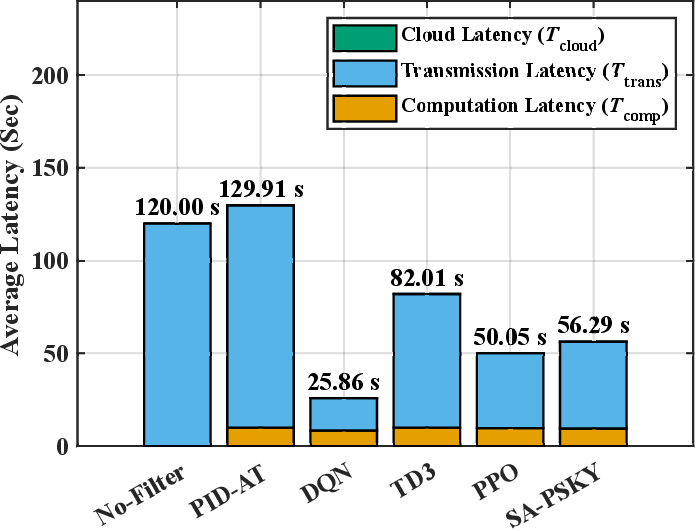}
    }\hfill
    \subfloat[Latency Distribution]{
        \label{fig:latency:b}
        \includegraphics[width=0.315\linewidth]{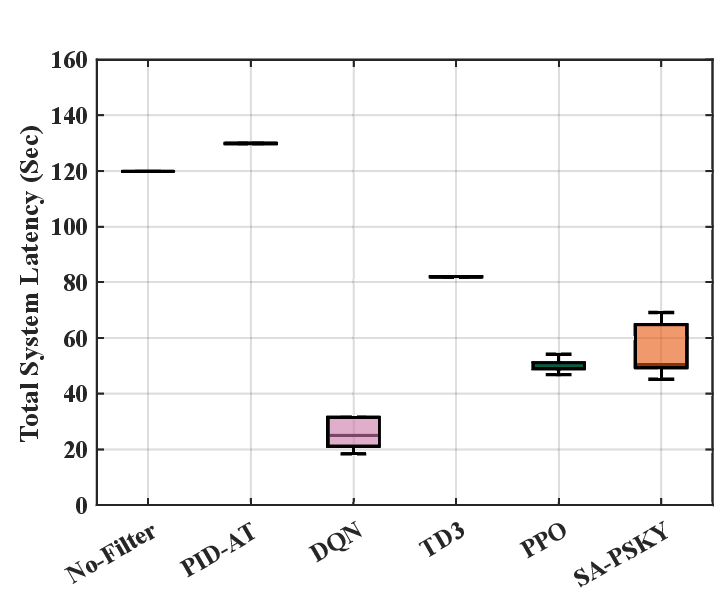}
    }\hfill
    \subfloat[Latency CDF]{
        \label{fig:latency:c}
        \includegraphics[width=0.315\linewidth]{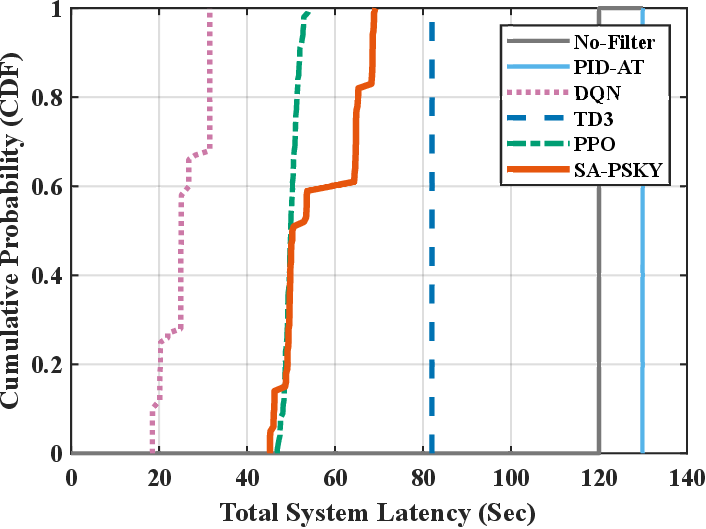}
    }
    \caption{Comprehensive latency evaluation of the proposed SA-PSKY against baseline algorithms. (a) Average system latency components comparison. (b) Distribution of total system latency over inference episodes. (c) Cumulative distribution function of total system latency.}
    \label{fig:latency}
\end{figure*}

System delay analysis (Fig.~\ref{fig:latency}) reveals the inherent trade-offs managed by SA-PSKY. The No-Filter baseline serves as the physical bound, where transmission bottlenecks dominate the latency profile. As shown in Fig.~\ref{fig:latency:a}, the No-Filter strategy averages 120.00 seconds. PID-AT fails to resolve this, reaching 129.91 seconds as its rigid heuristic logic ignores the coupling between computational cost and channel saturation. 

Compared to the No-Filter baseline, SA-PSKY allocates local filtering effort to remove unpromising candidates before transmission. By effectively pruning these candidates, our framework drastically reduces the payload size, cutting transmission delay by half. This demonstrates a core insight: in resource-constrained mobile edge environments, judiciously investing edge CPU cycles to filter data provides a highly effective approach for minimizing the overall end-to-end latency, as the massive savings in serialized transmission far outweigh the local processing costs. While DQN and PPO demonstrate lower absolute delays (25.86 and 50.05 seconds), they achieve this through aggressive, blind pruning that compromises query fidelity. SA-PSKY instead maintains a superior 56.29 seconds total latency by intelligently preserving global skyline candidates.

The adaptability insight is further highlighted in Fig.~\ref{fig:latency:b}. The rigid, zero-variance performance of No-Filter, TD3, and PID-AT reveals a fundamental inability to respond to traffic bursts. In contrast, the wider distribution of SA-PSKY reflects an active policy: the agent reserves extra computational cycles for rigorous filtering during high-traffic intervals to alleviate network congestion, while dynamically relaxing thresholds during idle periods to minimize processing overhead.

Finally, the \textit{Cumulative Distribution Function} (Fig.~\ref{fig:latency:c}) confirms QoS robustness. While No-Filter and PID-AT produce static, long-tail latency distributions, SA-PSKY exhibits a highly stable performance curve, with all observed inference latencies falling below 75 seconds. This validates that our agent does not merely optimize for the average case; rather, all observed latency values remain below 75 seconds, indicating consistent responsiveness under the tested spatio-temporal shifts.

\subsection{Ablation Study and Multi-Objective Pareto Analysis}
\label{subsec:ablation}

To dissect the individual contributions of the core architectural components within the SA-PSKY framework, we perform a systematic ablation study. We evaluate three distinct configurations: the full SA-PSKY framework with both PER and SAAW, a variant discarding PER, and a variant disabling SAAW. Fig.~\ref{fig:ablation} illustrates the training dynamics and objective space mapping to unveil the underlying policy trade-offs.

\subsubsection{Algorithmic Dexterity and Convergence Dynamics}
\begin{figure}[!t]
    \centering
    \subfloat[Total System Reward $U_{\rm sys}$]{
        \label{fig:ablation:a}
        \includegraphics[width=0.48\linewidth]{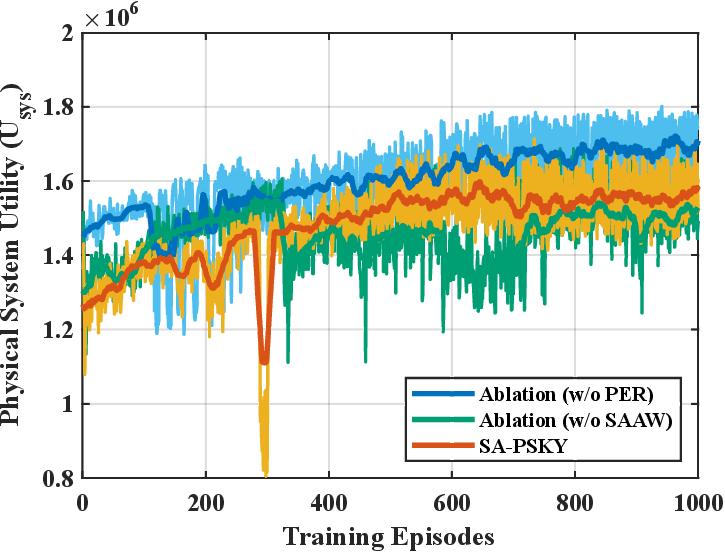}
    }%
    \subfloat[Data Selectivity $\Omega_t$]{
        \label{fig:ablation:b}
        \includegraphics[width=0.48\linewidth]{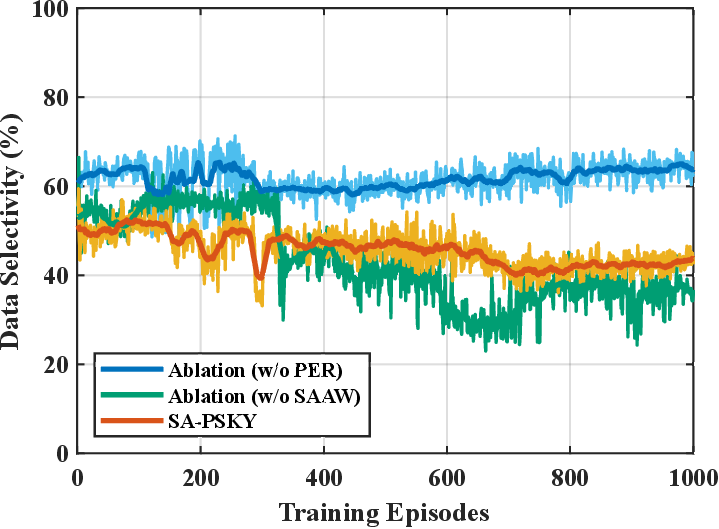}
    }\\
    \subfloat[Hit Rate $\tilde{\mathcal{U}}_{t}$]{
        \label{fig:ablation:c}
        \includegraphics[width=0.48\linewidth]{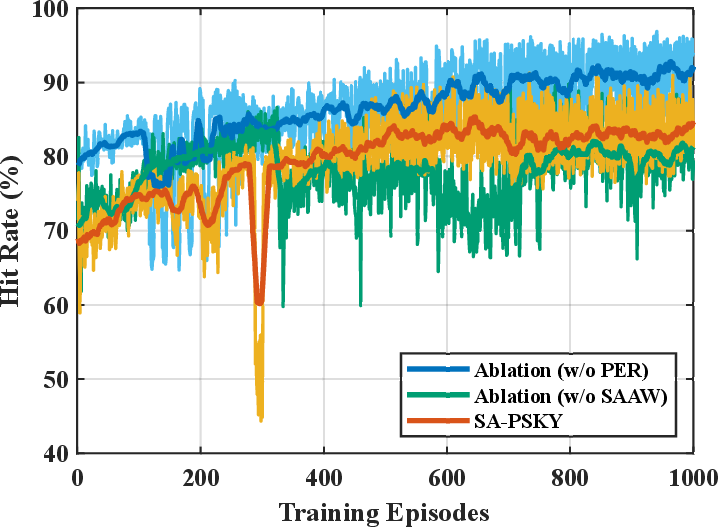}
    }%
    \subfloat[Pareto-Efficiency Analysis]{
        \label{fig:ablation:d}
        \includegraphics[width=0.48\linewidth]{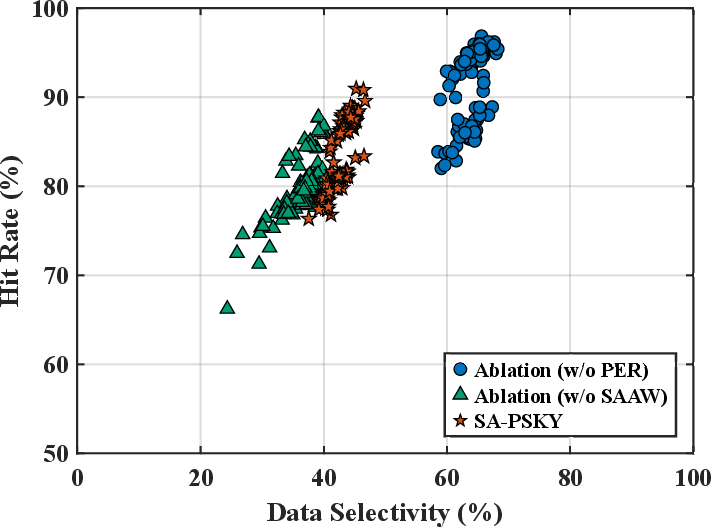}
    }
    \caption{Performance evaluation of the SA-PSKY architectural components. The training trajectories in (a)-(c) highlight the impact of PER and SAAW on learning stability, while the scatter plot in (d) indicates a superior trade-off of the full framework in the objective space.}
    \label{fig:ablation}
\end{figure}

The convergence characteristics illustrated in Fig.~\ref{fig:ablation:a} through~\ref{fig:ablation:c} reveal the structural necessity of each module. As shown in Fig.~\ref{fig:ablation:a} and Fig.~\ref{fig:ablation:c}, the configuration without PER exhibits a deceptively high query hit rate and total system reward. However, this superficial advantage is undermined by its data selectivity trajectory in Fig.~\ref{fig:ablation:b}, which consistently exceeds 60\%. Without the targeted sampling of high-error, boundary-violating transitions provided by PER, the deterministic agent stagnates in a lazy local optimum. It essentially transmits excessive data to guarantee query hits, failing to synthesize a precise filtering policy and effectively ignoring the uplink bandwidth constraints.

Conversely, the model disabling SAAW demonstrates the poorest overall system utility. In IoE environments characterized by spatio-temporal heterogeneity, relying on a static penalty weight vector renders the agent incapable of responding to fluctuating network congestion. To avoid severe queue overflow penalties, this ablated agent exhibits an overly conservative behavior, suppressing selectivity to near 30\%. While this aggressive cutoff preserves bandwidth, it blindly discards critical semantic information, causing the query hit rate to plummet.

The full SA-PSKY framework leverages SAAW to dynamically rebalance optimization priorities while utilizing PER to learn efficiently from critical system states. It successfully maintains a stable selectivity between 40\% and 45\% while restoring the hit rate to a competitive 85\%, thereby confirming its algorithmic dexterity in preserving system stability across complex traffic distributions.

\subsubsection{Empirical Pareto-Efficiency Analysis in the Objective Space}
The critical justification for the full SA-PSKY framework is established in the multi-objective scatter plot depicted in Fig.~\ref{fig:ablation:d}, which maps the joint distribution of data selectivity and query hit rate during the evaluation phase. In this objective space, the favorable operating regime lies strictly toward the top-left corner, representing maximum query fidelity at minimum bandwidth cost.

The visualization confirms that SAAW acts as a critical catalyst, driving the agent to move toward a Pareto-efficient operating region. The ablation model without PER is severely clustered in the resource-inefficient right region, while the model without SAAW sinks into the low-accuracy bottom-left region. The full SA-PSKY agent strategically bridges this performance gap. By effectively managing intrinsic spatio-temporal dynamics and workload variances through dynamic weight adjustments, it occupies a favorable Pareto-efficient region. It successfully secures a high hit rate while strictly confining data transmission volume, thereby indicating a superior trade-off compared to the rigid baseline variants and achieving a favorable operating balance for resource-constrained mobile edge environments.

\subsection{Zero-Shot Robustness to Mobility-Induced Uncertainty Shifts}
\label{subsec:generalization}

To further validate the practical applicability of the proposed SA-PSKY framework, we conduct a rigorous zero-shot generalization analysis under unseen mobility-induced uncertainty shifts. In real-world mobile edge networks, such as connected vehicle systems and mobile crowdsensing deployments, the physical mobility intensity of sensing nodes directly dictates the uncertainty footprint of the collected data. Specifically, higher node mobility and rapid spatial transitions exacerbate tracking and sensing ambiguity, which are abstracted in our model as a larger number of discrete instances, denoted as $m$, required to accurately capture the probability distribution of an uncertain object. A robust mobile edge intelligence framework must autonomously adapt to these massive-scale, mobility-induced variations without requiring continuous policy retraining.

\begin{figure}[!t]
    \centering
    \subfloat[Total System Reward $U_{\rm sys}$]{
        \label{fig:scalability:a}
        \includegraphics[width=0.48\linewidth]{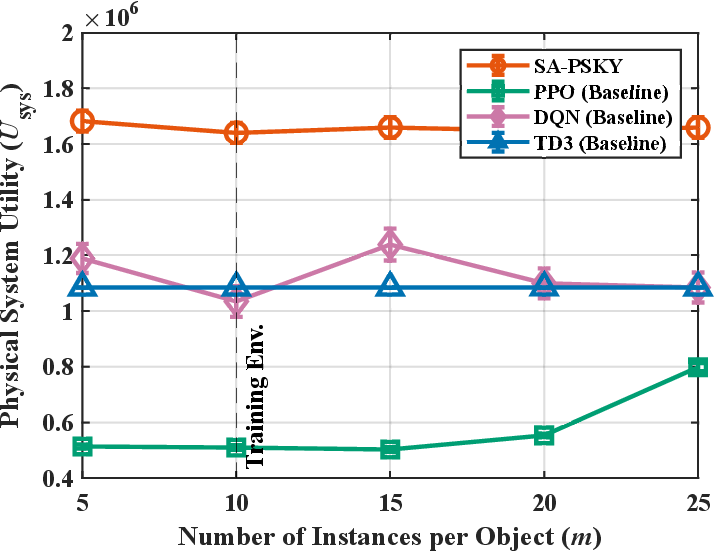}
    }%
    \subfloat[Data Selectivity $\Omega_t$]{
        \label{fig:scalability:b}
        \includegraphics[width=0.48\linewidth]{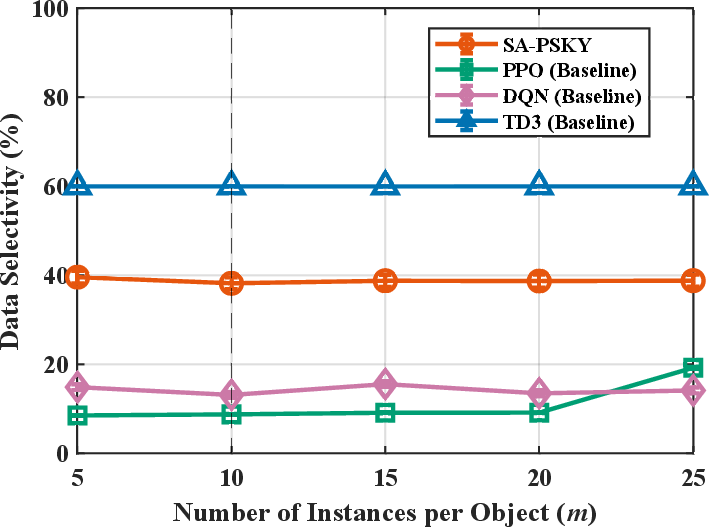}
    }\\
    \subfloat[Hit Rate $\tilde{\mathcal{U}}_{t}$]{
        \label{fig:scalability:c}
        \includegraphics[width=0.48\linewidth]{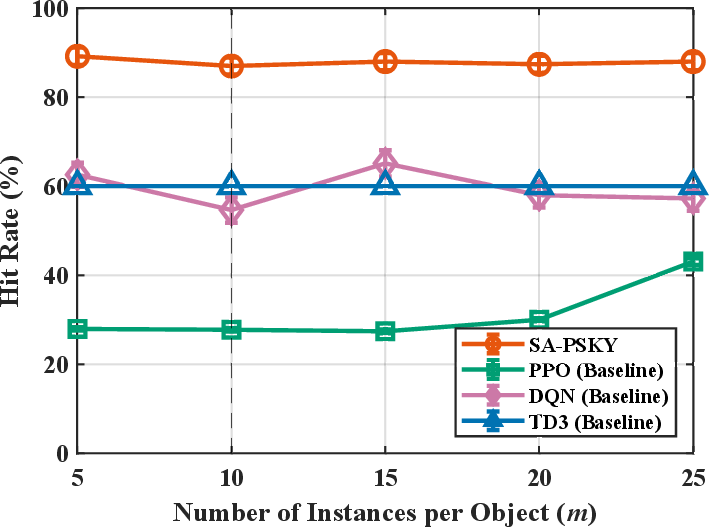}
    }%
    \subfloat[Objective Space Trajectory]{
        \label{fig:scalability:d}
        \includegraphics[width=0.47\linewidth]{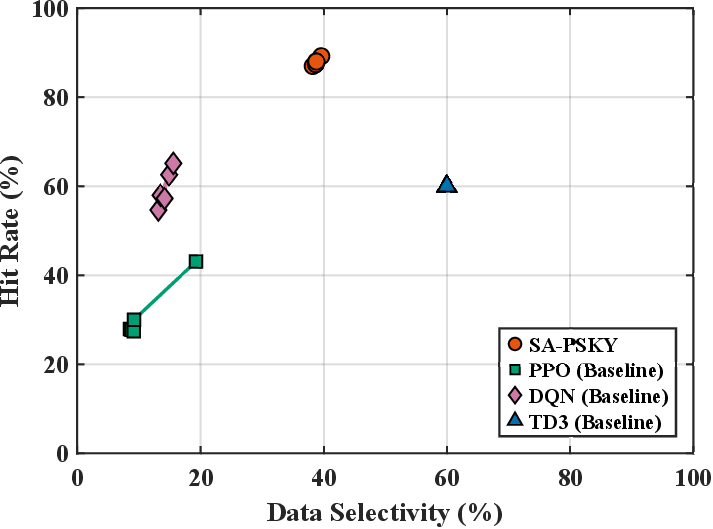}
    }
    \caption{Mobility robustness and zero-shot generalization analysis under varying mobility-induced uncertainty levels, parameterized by the number of instances per object ($m$). All models are trained exclusively at $m=10$. (a) Total system reward. (b) Data selectivity. (c) Query hit rate. (d) Pareto trajectory in the objective space.}
    \label{fig:scalability}
\end{figure}

In this experiment, all evaluated models are exclusively trained in a baseline environment configured with a moderate mobility profile ($m=10$). Subsequently, we freeze the network weights and directly deploy the agents in unseen environments where the mobility intensity scales dramatically, expanding $m$ from 5 (low mobility) to 25 (high mobility). Fig.~\ref{fig:scalability} illustrates the performance degradation and the objective space trajectories of SA-PSKY alongside the baseline methods under this zero-shot transfer configuration.

As the mobility intensity $m$ increases, the dimensionality of the state space and the probability overlap among objects expand significantly, revealing that baseline methods fail to handle distribution shifts and diverge into suboptimal policies. On one end of the spectrum, the deterministic TD3 baseline exhibits a completely rigid and unyielding policy under zero-shot evaluation. Its data selectivity is locked at exactly 60 percent and its hit rate stagnates at roughly 60 percent, regardless of the varying data densities. This indicates that TD3 fails to extract meaningful spatial distribution features during transfer, settling into a rigid local optimum that compromises both bandwidth and accuracy. On the opposite extreme, the stochastic PPO baseline falls into a severe under-selection trap. Without the continuous stochastic exploration present during the training phase, PPO exhibits severe degradation under the deterministic zero-shot evaluation setting. As shown in Fig.~\ref{fig:scalability:b}, PPO severely drops its data selectivity to below 10 percent for most test cases, only slightly recovering under extreme mobility ($m=25$). This severe over-suppression blocks critical skyline features, causing the query hit rate in Fig.~\ref{fig:scalability:c} to plummet to approximately 30 percent while accumulating massive negative rewards. Similarly, the discrete action space of DQN confines it to a conservative local optimum. While it manages a hit rate between 55 percent and 65 percent, its selectivity remains overly restricted below 20 percent, and it exhibits unstable fluctuations across different mobility profiles, indicating its limited ability to adapt smoothly to varying mobility-induced uncertainty levels.

Conversely, SA-PSKY serves as the effective algorithmic compromise, successfully navigating the continuous Pareto cliff between data flood and data starvation. Remarkably, although the mobility intensity parameter $m$ and the instantaneous computational load are not explicitly observable within the agent's state vector $s_t$, SA-PSKY still generalizes effectively. By anchoring its policy strictly on spatial distribution features ($\mathcal{D}_t$) and dynamically recalibrating the filtering thresholds via the SAAW module, it maintains a remarkably stable selectivity bound ($\Omega_t$) at approximately 40 percent across all unknown mobility profiles. Simultaneously, it secures a dominant and stable query hit rate approaching 90 percent. This stable and aggressive filtering implicitly minimizes the pruning efficiency factor $\Phi(\alpha)$, thereby absorbing the severe $O(m^2)$ computational complexity and preventing local CPU saturation without requiring explicitly engineered state variables. This deterministic stability ensures that SA-PSKY comprehensively avoids both the under-selection trap of PPO and the rigid sub-optimality of TD3, confirming its robustness in highly dynamic mobile environments.

\subsection{Summary of Experimental Findings}

The comprehensive evaluation presented in this section validates the efficacy, architectural necessity, and robust mobility adaptation of the SA-PSKY framework under diverse mobile network conditions. The key findings are summarized as follows:

\begin{enumerate}
    \item \textbf{Algorithmic Dominance}: Our comparative evaluation confirms that SA-PSKY significantly outperforms existing DRL baselines. By integrating a PER buffer to improve learning robustness and SAAW to adaptively rebalance optimization priorities, SA-PSKY avoids the policy collapse observed in TD3 and the suboptimal conservatism characteristic of DQN and PPO. 

    \item \textbf{Structural Necessity}: The ablation study elucidates the indispensable role of the core architectural components. The ablation study shows that PER substantially improves policy stability near strict boundary constraints. Furthermore, SAAW provides the necessary algorithmic dexterity to navigate the objective space, transforming the agent from a resource-inefficient policy into an intelligent decision-maker capable of aggressive bandwidth compression.

    \item \textbf{Mobility Robustness}: The zero-shot generalization analysis underscores the robustness and adaptability of SA-PSKY in dynamic scenarios. While baseline methods exhibit extreme brittleness under mobility-induced distribution shifts, manifesting as either rigid sub-optimality or severe data-starvation traps, SA-PSKY maintains consistent performance across varying mobility intensities. Its ability to dynamically recalibrate filtering thresholds without requiring policy retraining confirms that the framework effectively captures the underlying computation-communication trade-off of edge filtering, providing a resilient foundation for real-time analytics in complex, heterogeneous mobile environments.
\end{enumerate}

These findings collectively demonstrate that SA-PSKY offers a robust, Pareto-efficient, and intelligent solution for processing uncertain data streams, yielding consistent performance gains for next-generation resource-constrained mobile edge-cloud ecosystems.


\section{Conclusion}
\label{sec:conclusion}

This paper has presented SA-PSKY, a self-adaptive framework designed to resolve the critical tension between computational overhead, communication bandwidth, and query fidelity in distributed probabilistic skyline analytics. By formulating threshold selection as a continuous control task and integrating the SAAW module alongside the PER buffer, SA-PSKY empowers mobile edge nodes to learn an adaptive Pareto-efficient operating point under varying mobile resource dynamics, rather than trivially minimizing latency by aggressively discarding data.

While conventional baseline methods often succumb to policy collapse or resource-inefficient local optima under volatile data streams, our analysis confirms that SA-PSKY achieves robust zero-shot generalization and stability. Ultimately, this framework demonstrates that performing local filtering at the edge while retaining global verification at the cloud, governed by an intelligent and state-aware agent, provides a practical architectural framework for adaptive probabilistic skyline analytics in complex mobile and heterogeneous edge-cloud environments.

Finally, we acknowledge certain limitations in the current framework. The system evaluates mobility robustness primarily through mobility-induced workload and spatial distribution dynamics, utilizing an abstracted wireless communication model and a simplified M/M/1 queue. It does not explicitly model full physical layer (PHY) and medium access control (MAC) channel fading or discrete handover mechanisms. Future research will focus on integrating real-world vehicular mobility traces and explicit wireless channel simulations to further extend the framework's applicability under extreme network conditions.


\section*{Acknowledgment}
The author would like to thank Wei-Hong Chen for his assistance in the preliminary experimental data collection during his Master's studies at Feng Chia University. The current framework represents a fundamental architectural departure from that study, evolving from a discrete DQN-based approach to a continuous-control DDPG paradigm, and incorporating multi-objective reward shaping with state-aware adaptive weighting.


\bibliographystyle{IEEEtran}
\bibliography{IEEEabrv,reference}

@STRING{IEEE_J_BD         = "{IEEE} Trans. Big Data"}

@STRING{IEEE_J_KDE        = "{IEEE} Trans. Knowl. Data Eng."}

@STRING{IEEE_J_IOT        = "{IEEE} Internet Things J."}

@STRING{IEEE_J_CC         = "{IEEE} Trans. on Cloud Comput."}

@STRING{IEEE_J_C          = "{IEEE} Trans. Comput."}

@STRING{IEEE_J_ETC        = "{IEEE} Trans. Emerg. Topics Comput."}

@STRING{IEEE_J_MC         = "{IEEE} Trans. Mobile Comput."}

@STRING{IEEE_J_PDS        = "{IEEE} Trans. Parallel Distrib. Syst."}

@STRING{IEEE_J_SC         = "{IEEE} Trans. Serv. Comput."}

@STRING{IEEE_O_CSTO       = "{IEEE} Commun. Surveys Tuts."}

@article{11244906,
  author  = {Adil, Muhammad and Qiu, Tie and Zhou, Xiaobo and Javeed, Danish and Cao, Zhenrui and Oliver Wu, Dapeng},
  journal = IEEE_O_CSTO,
  title   = {Integrated {5G} and Time Sensitive Networking for Emerging Applications: A Survey of Advancements, Challenges, and Future Directions},
  year    = {2026},
  volume  = {28},
  number  = {},
  pages   = {4016-4050},
  doi     = {10.1109/COMST.2025.3632286}
}

@article{11074426,
  author  = {He, Qiang and Lin, Jinqiu and Fang, Hui and Wang, Xingwei and Huang, Min and Yi, Xiushuang and Yu, Keping},
  journal = IEEE_J_SC,
  title   = {Integrating {IoT} and {6G}: Applications of Edge Intelligence, Challenges, and Future Directions},
  year    = {2025},
  volume  = {18},
  number  = {4},
  pages   = {2471-2488},
  month   = {Jul.-Aug.},
  doi     = {10.1109/TSC.2025.3586152}
}

@article{10545344,
  author  = {Yang, Ning and Chen, Shuo and Zhang, Haijun and Berry, Randall},
  journal = IEEE_O_CSTO,
  title   = {Beyond the Edge: An Advanced Exploration of Reinforcement Learning for Mobile Edge Computing, Its Applications, and Future Research Trajectories},
  year    = {2025},
  volume  = {27},
  number  = {1},
  pages   = {546-594},
  month   = feb,
  doi     = {10.1109/COMST.2024.3405075}
}

@article{11098465,
  author  = {Liu, Zhiyan and Chen, Xu and Wu, Hai and Wang, Zhanwei and Chen, Xianhao and Niyato, Dusit and Huang, Kaibin},
  journal = IEEE_O_CSTO,
  title   = {Integrated Sensing and Edge {AI}: Realizing Intelligent Perception in {6G}},
  year    = {2026},
  volume  = {28},
  number  = {},
  pages   = {2725-2770},
  doi     = {10.1109/COMST.2025.3592989}
}

@article{9374102,
  author  = {Luo, Quyuan and Li, Changle and Luan, Tom H. and Shi, Weisong},
  journal = IEEE_J_SC,
  title   = {Minimizing the Delay and Cost of Computation Offloading for Vehicular Edge Computing},
  year    = {2022},
  volume  = {15},
  number  = {5},
  pages   = {2897-2909},
  month   = {Sept.-Oct.},
  doi     = {10.1109/TSC.2021.3064579}
}

@article{9633191,
  author  = {Pan, Lei and Liu, Xiao and Jia, Zhaohong and Xu, Jia and Li, Xuejun},
  journal = IEEE_J_CC,
  title   = {A Multi-Objective Clustering Evolutionary Algorithm for Multi-Workflow Computation Offloading in Mobile Edge Computing},
  year    = {2023},
  volume  = {11},
  number  = {2},
  pages   = {1334-1351},
  month   = {Apr.-Jun.},
  doi     = {10.1109/TCC.2021.3132175}
}

@article{10595132,
  author  = {Li, Ruoyu and Li, Qing and Zou, Qingsong and Zhao, Dan and Zeng, Xiangyi and Huang, Yucheng and Jiang, Yong and Lyu, Feng and Ormazabal, Gaston and Singh, Aman and Schulzrinne, Henning},
  journal = IEEE_J_MC,
  title   = {{IoTGemini}: Modeling {IoT} Network Behaviors for Synthetic Traffic Generation},
  year    = {2024},
  volume  = {23},
  number  = {12},
  pages   = {13240-13257},
  month   = dec,
  doi     = {10.1109/TMC.2024.3426600}
}

@article{10175534,
  author  = {Hussein, Dana Haj and Ibnkahla, Mohamed},
  journal = IEEE_J_IOT,
  title   = {A Novel Mathematical Framework for Modeling Application-Specific IoT Traffic},
  year    = {2024},
  volume  = {11},
  number  = {2},
  pages   = {2364-2381},
  month   = jan,
  doi     = {10.1109/JIOT.2023.3293028}
}

@article{10597395,
  author  = {Le, Mai and Huynh-The, Thien and Do-Duy, Tan and Vu, Thai-Hoc and Hwang, Won-Joo and Pham, Quoc-Viet},
  journal = IEEE_O_CSTO,
  title   = {Applications of Distributed Machine Learning for the {Internet-of-Things}: A Comprehensive Survey},
  year    = {2025},
  volume  = {27},
  number  = {2},
  pages   = {1053-1100},
  month   = apr,
  doi     = {10.1109/COMST.2024.3427324}
}

@inproceedings{10.5555/1325851.1325858,
  author    = {Pei, Jian and Jiang, Bin and Lin, Xuemin and Yuan, Yidong},
  title     = {Probabilistic skylines on uncertain data},
  year      = {2007},
  booktitle = {The 33rd International Conference on Very Large Data Bases},
  address   = {Vienna, Austria}
}

@article{LIU201540,
  title   = {An Effective Probabilistic Skyline Query Process on Uncertain Data Streams},
  journal = {Procedia Computer Science},
  volume  = {63},
  pages   = {40-47},
  year    = {2015},
  month   = sep,
  doi     = {10.1016/j.procs.2015.08.310},
  author  = {Chuan-Ming Liu and Syuan-Wei Tang}
}

@article{10298037,
  author  = {Lai, Chuan-Chi and Lin, Hsuan-Yu and Liu, Chuan-Ming},
  journal = IEEE_J_ETC,
  title   = {Distributed Indexing Schemes for K-Dominant Skyline Analytics on Uncertain Edge-{IoT} Data},
  year    = {2024},
  volume  = {12},
  number  = {3},
  pages   = {878-890},
  month   = {Jul.-Sept.},
  doi     = {10.1109/TETC.2023.3326295}
}

@article{9354847,
  author  = {Zhang, Xichen and Lu, Rongxing and Shao, Jun and Zhu, Hui and Ghorbani, Ali A.},
  journal = IEEE_J_IOT,
  title   = {Continuous Probabilistic Skyline Query for Secure Worker Selection in Mobile Crowdsensing},
  year    = {2021},
  volume  = {8},
  number  = {14},
  pages   = {11758-11772},
  month   = jul,
  doi     = {10.1109/JIOT.2021.3059637}
}

@article{10356753,
  author  = {Liang, Huanghuang and Zhang, Zheng and Hu, Chuang and Gong, Yili and Cheng, Dazhao},
  journal = IEEE_J_BD,
  title   = {A Survey on Spatio-Temporal Big Data Analytics Ecosystem: Resource Management, Processing Platform, and Applications},
  year    = {2024},
  volume  = {10},
  number  = {2},
  pages   = {174-193},
  month   = apr,
  doi     = {10.1109/TBDATA.2023.3342619}
}

@inproceedings{914855,
  author    = {Borzsony, S. and Kossmann, D. and Stocker, K.},
  booktitle = {Proceedings of 17th International Conference on Data Engineering},
  title     = {The Skyline operator},
  year      = {2001},
  volume    = {},
  number    = {},
  pages     = {421-430},
  doi       = {10.1109/ICDE.2001.914855}
}

@article{5394984,
  author  = {Lian, Xiang and Chen, Lei},
  journal = IEEE_J_KDE,
  title   = {Ranked Query Processing in Uncertain Databases},
  year    = {2010},
  volume  = {22},
  number  = {3},
  pages   = {420-436},
  month   = mar,
  doi     = {10.1109/TKDE.2009.112}
}

@inproceedings{4221738,
  author    = {Soliman, Mohamed A. and Ilyas, Ihab F. and Chen-Chuan Chang, Kevin},
  booktitle = {IEEE 23rd International Conference on Data Engineering (ICDE)},
  title     = {Top-k Query Processing in Uncertain Databases},
  year      = {2007},
  doi       = {10.1109/ICDE.2007.367935},
  address   = {Istanbul, Turkey}
}

@article{10.1145/1670243.1670246,
  author    = {Lian, Xiang and Chen, Lei},
  title     = {Reverse skyline search in uncertain databases},
  year      = {2008},
  volume    = {35},
  number    = {1},
  doi       = {10.1145/1670243.1670246},
  journal   = {ACM Transactions on Database Systems},
  month     = feb,
  articleno = {3},
  numpages  = {49}
}

@inproceedings{1410162,
  author    = {Xuemin Lin and Yidong Yuan and Wei Wang and Hongjun Lu},
  booktitle = {21st International Conference on Data Engineering (ICDE)},
  title     = {Stabbing the sky: efficient skyline computation over sliding windows},
  year      = {2005},
  doi       = {10.1109/ICDE.2005.137},
  address   = {Tokyo, Japan}
}

@article{8731646,
  author  = {Lai, Chuan-Chi and Wang, Tien-Chun and Liu, Chuan-Ming and Wang, Li-Chun},
  journal = IEEE_J_IOT,
  title   = {Probabilistic Top-${k}$  Dominating Query Monitoring Over Multiple Uncertain {IoT} Data Streams in Edge Computing Environments},
  year    = {2019},
  volume  = {6},
  number  = {5},
  pages   = {8563-8576},
  month   = oct,
  doi     = {10.1109/JIOT.2019.2920908}
}

@book{2010Neely,
  author    = {Michael J. Neely},
  title     = {Stochastic Network Optimization with Application to Communication and Queueing Systems},
  series    = {Synthesis Lectures on Communication Networks},
  publisher = {Morgan {\&} Claypool Publishers},
  year      = {2010},
  doi       = {10.2200/S00271ED1V01Y201006CNT007},
  isbn      = {978-3-031-79994-5}
}

@article{7488250,
  author  = {Shi, Weisong and Cao, Jie and Zhang, Quan and Li, Youhuizi and Xu, Lanyu},
  journal = IEEE_J_IOT,
  title   = {Edge Computing: Vision and Challenges},
  year    = {2016},
  volume  = {3},
  number  = {5},
  pages   = {637-646},
  month   = oct,
  doi     = {10.1109/JIOT.2016.2579198}
}

@article{Gao2025,
  author  = {Xiangyu Gao and Xingxing Xiao and Xiao Pan and Dongjing Miao and Jianzhong Li},
  title   = {Efficient Algorithms for Uncertain Restricted Skyline Query Processing},
  journal = {The VLDB Journal},
  volume  = {34},
  number  = {4},
  pages   = {46},
  year    = {2025},
  month   = may,
  doi     = {10.1007/s00778-025-00925-9}
}

@inproceedings{7379908,
  author    = {Santoso, Bagus Jati and Chiu, Ge-Ming and Mumpuni, Retno},
  booktitle = {International Conference on Information \& Communication Technology and Systems (ICTS)},
  title     = {An efficient grid-based framework for answering tolerance-based skyline queries},
  year      = {2015},
  address   = {Surabaya, Indonesia},
  doi       = {10.1109/ICTS.2015.7379908}
}

@inproceedings{10.1007/11687238_10,
  author    = {Wu, Ping
               and Zhang, Caijie
               and Feng, Ying
               and Zhao, Ben Y.
               and Agrawal, Divyakant
               and El Abbadi, Amr},
  editor    = {Ioannidis, Yannis
               and Scholl, Marc H.
               and Schmidt, Joachim W.
               and Matthes, Florian
               and Hatzopoulos, Mike
               and Boehm, Klemens
               and Kemper, Alfons
               and Grust, Torsten
               and Boehm, Christian},
  title     = {Parallelizing Skyline Queries for Scalable Distribution},
  booktitle = {Advances in Database Technology - EDBT 2006},
  year      = {2006},
  publisher = {Springer Berlin Heidelberg},
  address   = {Berlin, Heidelberg},
  pages     = {112--130}
}

@article{4731252,
  author  = {Cui, Bin and Chen, Lijiang and Xu, Linhao and Lu, Hua and Song, Guojie and Xu, Quanqing},
  journal = IEEE_J_KDE,
  title   = {Efficient Skyline Computation in Structured Peer-to-Peer Systems},
  year    = {2009},
  volume  = {21},
  number  = {7},
  pages   = {1059-1072},
  month   = jul,
  doi     = {10.1109/TKDE.2008.235}
}

@article{SU2010680,
  title   = {Efficient skyline query processing in wireless sensor networks},
  journal = {Journal of Parallel and Distributed Computing},
  volume  = {70},
  number  = {6},
  pages   = {680-698},
  year    = {2010},
  issn    = {0743-7315},
  doi     = {10.1016/j.jpdc.2010.01.001},
  month   = jun,
  author  = {I-Fang Su and Yu-Chi Chung and Chiang Lee and Yi-Ying Lin}
}

@article{9714833,
  author  = {Kuo, Ai-Te and Chen, Haiquan and Tang, Liang and Ku, Wei-Shinn and Qin, Xiao},
  journal = IEEE_J_KDE,
  title   = {{ProbSky}: Efficient Computation of Probabilistic Skyline Queries Over Distributed Data},
  year    = {2023},
  volume  = {35},
  number  = {5},
  pages   = {5173-5186},
  month   = may,
  doi     = {10.1109/TKDE.2022.3151740}
}

@inproceedings{10.1145/1376616.1376642,
  author    = {Vlachou, Akrivi and Doulkeridis, Christos and Kotidis, Yannis},
  title     = {Angle-based space partitioning for efficient parallel skyline computation},
  year      = {2008},
  doi       = {10.1145/1376616.1376642},
  booktitle = {The 2008 ACM SIGMOD International Conference on Management of Data},
  address   = {Vancouver, Canada}
}

@article{9186333,
  author  = {Wijayanto, Heri and Wang, Wenlu and Ku, Wei-Shinn and Chen, Arbee L.P.},
  journal = IEEE_J_KDE,
  title   = {{LShape} Partitioning: Parallel Skyline Query Processing Using {MapReduce}},
  year    = {2022},
  volume  = {34},
  number  = {07},
  issn    = {1558-2191},
  pages   = {3363-3376},
  doi     = {10.1109/TKDE.2020.3021470},
  month   = jul
}

@article{NGUYEN2024103884,
  title   = {Exploring the integration of edge computing and blockchain {IoT}: Principles, architectures, security, and applications},
  journal = {Journal of Network and Computer Applications},
  volume  = {226},
  pages   = {103884},
  year    = {2024},
  issn    = {1084-8045},
  doi     = {10.1016/j.jnca.2024.103884},
  author  = {Tri Nguyen and Huong Nguyen and Tuan {Nguyen Gia}}
}

@article{9863881,
  author  = {Kong, Xiangjie and Wu, Yuhan and Wang, Hui and Xia, Feng},
  journal = IEEE_J_IOT,
  title   = {Edge Computing for {Internet of Everything}: A Survey},
  year    = {2022},
  volume  = {9},
  number  = {23},
  pages   = {23472-23485},
  month   = dec,
  doi     = {10.1109/JIOT.2022.3200431}
}

@article{9723570,
  author  = {Cai, Qingpeng and Cui, Can and Xiong, Yiyuan and Wang, Wei and Xie, Zhongle and Zhang, Meihui},
  journal = IEEE_J_KDE,
  title   = {A Survey on Deep Reinforcement Learning for Data Processing and Analytics},
  year    = {2023},
  volume  = {35},
  number  = {5},
  pages   = {4446-4465},
  month   = may,
  doi     = {10.1109/TKDE.2022.3155196}
}

@article{10316604,
  author  = {Xu, Zichuan and Xu, Guangyuan and Wang, Hao and Liang, Weifa and Xia, Qiufen and Wang, Shangguang},
  journal = IEEE_J_PDS,
  title   = {Enabling Streaming Analytics in Satellite Edge Computing via Timely Evaluation of Big Data Queries},
  year    = {2024},
  volume  = {35},
  number  = {1},
  pages   = {105-122},
  month   = jan,
  doi     = {10.1109/TPDS.2023.3332333}
}

@article{8302507,
  author  = {Tang, Mingjie and Yu, Yongyang and Aref, Walid G. and Malluhi, Qutaibah M. and Ouzzani, Mourad},
  journal = IEEE_J_KDE,
  title   = {Efficient Parallel Skyline Query Processing for High-Dimensional Data},
  year    = {2018},
  volume  = {30},
  number  = {10},
  pages   = {1838-1851},
  month   = oct,
  doi     = {10.1109/TKDE.2018.2809598}
}

@inproceedings{Graham2023Edge,
  author    = {Graham, Jessica and Medico, Anthony and Dividino, Renata and De Grande, Robson E.},
  title     = {Edge Clustering and Communication Efficiency with {GNNs} in {Internet of Vehicles}},
  booktitle = {The 6th International Conference on Wireless, Intelligent and Distributed Environment for Communication (WIDECOM)},
  address   = {Ontario, Canada},
  year      = {2023}
}

@article{10308406,
  author  = {Cui, Yangguang and Zhang, Zhixing and Wang, Nuo and Li, Liying and Chang, Chunwei and Wei, Tongquan},
  journal = IEEE_J_C,
  title   = {User-Distribution-Aware Federated Learning for Efficient Communication and Fast Inference},
  year    = {2024},
  volume  = {73},
  number  = {4},
  pages   = {1004-1018},
  month   = apr,
  doi     = {10.1109/TC.2023.3327513}
}

@article{9944146,
  author  = {Mangiaracina, Giulia and Plebani, Pierluigi and Salnitri, Mattia and Vitali, Monica},
  journal = IEEE_J_CC,
  title   = {Efficient Data as a Service in Fog Computing: An Adaptive Multi-Agent Based Approach},
  year    = {2023},
  volume  = {11},
  number  = {3},
  pages   = {2646-2663},
  month   = {Jul.-Sept.},
  doi     = {10.1109/TCC.2022.3220811}
}

@article{11494288,
  author  = {Zhai, Xue and Pang, Shanchen and Qiao, Sibo and Yu, Shihang and Gui, Haiyuan},
  journal = IEEE_J_CC,
  title   = {{AMIMSO}: Adaptive {Markov} Inspired Mutation Swarm Optimization for Multi-Objective Task Scheduling in {End-Edge-Cloud} Environments},
  year    = {2026},
  volume  = {},
  number  = {},
  pages   = {1-16},
  note    = {Early Access},
  doi     = {10.1109/TCC.2026.3687347}
}

@inproceedings{per2016,
  author    = {Tom Schaul and
               John Quan and
               Ioannis Antonoglou and
               David Silver},
  title     = {Prioritized Experience Replay},
  booktitle = {The 4th International Conference on Learning Representations (ICLR)},
  year      = {2016},
  address   = {San Juan, Puerto Rico}
}

@article{mnih2015human,
  title     = {Human-level control through deep reinforcement learning},
  author    = {Mnih, Volodymyr and Kavukcuoglu, Koray and Silver, David and Rusu, Andrei A and Veness, Joel and Bellemare, Marc G and Graves, Alex and Riedmiller, Martin and Fidjeland, Andreas K and Ostrovski, Georg and others},
  journal   = {Nature},
  volume    = {518},
  number    = {7540},
  pages     = {529--533},
  year      = {2015},
  publisher = {Nature Publishing Group}
}

@inproceedings{lillicrap2016,
  title     = {Continuous control with deep reinforcement learning},
  author    = {Lillicrap, Timothy P and Hunt, Jonathan J and Pritzel, Alexander and Heess, Nicolas and Erez, Tom and Tassa, Yuval and Silver, David and Wierstra, Daan},
  booktitle = {The 4th International Conference on Learning Representations (ICLR)},
  year      = {2016},
  address   = {San Juan, Puerto Rico}
}

@article{10.1007/s00778-009-0162-1,
  author   = {Zhang, Wenjie and Lin, Xuemin and Zhang, Ying and Pei, Jian and Wang, Wei},
  title    = {Threshold-based probabilistic top-k dominating queries},
  year     = {2010},
  volume   = {19},
  number   = {2},
  doi      = {10.1007/s00778-009-0162-1},
  journal  = {The VLDB Journal},
  month    = apr,
  pages    = {283--305},
  numpages = {23}
}
\ifCLASSOPTIONcaptionsoff  \newpage \fi 

\begin{IEEEbiography}[{\includegraphics[width=1in,height=1.25in,clip,keepaspectratio]{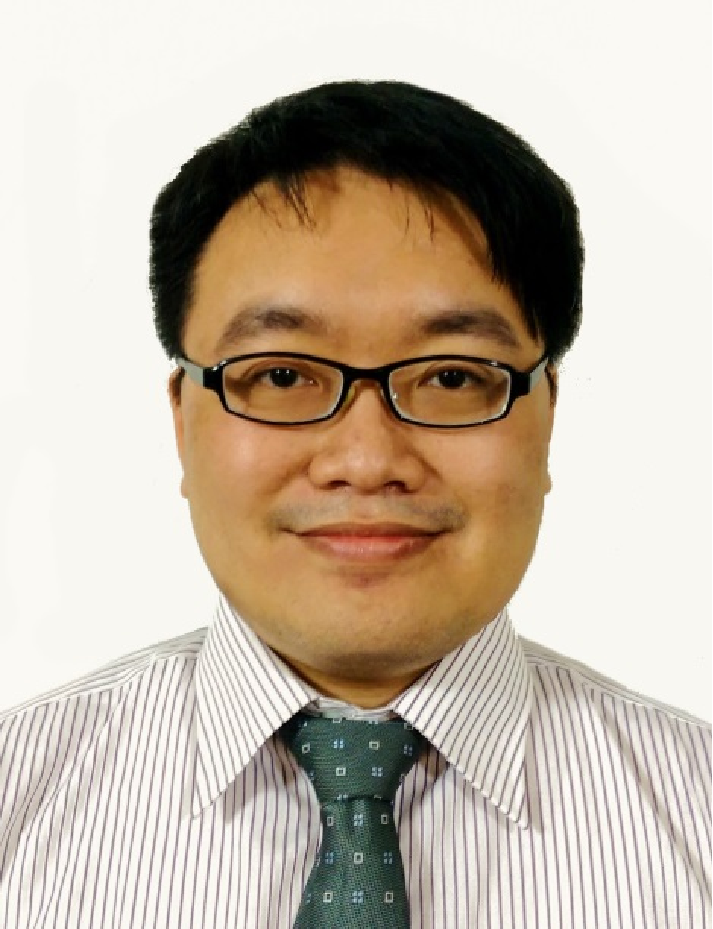}}]{Chuan-Chi Lai}
    (Member, IEEE) received the Ph.D. degree in Computer Science and Information Engineering from the National Taipei University of Technology, Taiwan, in 2017. He held research and faculty positions at National Chiao Tung University and Feng Chia University prior to his current role. Since 2024, he has been an Assistant Professor with the Department of Communications Engineering, National Chung Cheng University, Chiayi, Taiwan. His research interests include mobile edge computing, UAV networks, and AI for wireless communications. Dr. Lai was a recipient of the Postdoctoral Researcher Academic Research Award from the NSTC, Taiwan, in 2019, and Best Paper Awards at WOCC (2018, 2021) and ICUFN (2015).
\end{IEEEbiography}

\end{document}